\documentclass[tighten,iop]{emulateapj}

\usepackage[english]{babel}
\usepackage{amsmath}
\usepackage{amsfonts}
\usepackage{amssymb}
\usepackage{graphicx}
\usepackage{color}

\usepackage{natbib}

\shorttitle{Relativistic Explosion Models of Core-Collapse Supernovae}
\shortauthors{M\"uller, Janka \& Marek}

\begin{document}

\title{A New Multi-Dimensional General Relativistic Neutrino
Hydrodynamics Code for Core-Collapse Supernovae \\
II. Relativistic Explosion Models of Core-Collapse Supernovae}

\author{Bernhard~M\"uller, Hans-Thomas~Janka, Andreas~Marek}

\affil{Max-Planck-Institut f\"ur Astrophysik,                                                 \
       Karl-Schwarzschild-Str. 1, D-85748 Garching, Germany;                                  \
       bjmuellr@mpa-garching.mpg.de, thj@mpa-garching.mpg.de}

\begin{abstract}
We present the first two-dimensional general relativistic (GR)
simulations of stellar core collapse and explosion with the
\textsc{CoCoNuT }hydrodynamics code in combination with the
\textsc{Vertex} solver for energy-dependent, three-flavor neutrino
transport, using the extended conformal flatness condition for
approximating the spacetime metric and a ray-by-ray-plus ansatz to
tackle the multi-dimensionality of the transport. For both of the
investigated $11.2$ and $15\, M_\odot$ progenitors we obtain
successful, though seemingly marginal, neutrino-driven supernova
explosions. This outcome and the time evolution of the models
basically agree with results previously obtained with the
\textsc{Prometheus} hydro solver including an approximative treatment
of relativistic effects by a modified Newtonian potential. However, GR
models exhibit subtle differences in the neutrinospheric conditions
compared to Newtonian and pseudo-Newtonian simulations. These
differences lead to significantly higher luminosities and mean
energies of the radiated electron neutrinos and antineutrinos and
therefore to larger energy-deposition rates and heating efficiencies
in the gain layer with favorable consequences for strong nonradial
mass motions and ultimately for an explosion.
Moreover, energy
transfer to the stellar medium around the neutrinospheres through
nucleon recoil in scattering reactions of heavy-lepton neutrinos also
enhances the mentioned effects.
Together with previous
pseudo-Newtonian models the presented relativistic calculations
suggest that the treatment of gravity and energy-exchanging neutrino
interactions can make differences of even 50--100\% in some quantities
and is likely to contribute to a finally successful explosion
mechanism on no minor level than hydrodynamical differences between
different dimensions.
\end{abstract}


\section{Introduction}
\label{sec:intro}

More than 40 years after the first attempts by \citet{colgate_66}, the
quest for the supernova explosion mechanism is still ongoing. Since
the solution now seems (once again) within reach as several groups
have come up with successful explosion models
\citep{marek_09,suwa_10,takiwaki_12,bruenn_09,burrows_06,dessart_07_a},
the demand for accurate predictions of the neutrino and gravitational
wave signals and of the nucleosynthetic yields becomes more urgent and
naturally requires increased levels of sophistication in numerical
core-collapse supernova simulations.

Currently, there is a broad consensus that multi-dimensional
hydrodynamical instabilities are one of the pivotal elements of the
supernova problem. It has long been recognized that convection in the
hot-bubble region serves to increase the efficiency of neutrino
heating behind the shock
\citep{herant_92,herant_94,burrows_95,janka_96,mueller_97}, and that
another more recently discovered large-scale instability, the standing
accretion-shock instability (``SASI'',
\citealp{blondin_03,blondin_06,foglizzo_06,ohnishi_06,
  foglizzo_07,scheck_08,iwakami_08,iwakami_09,fernandez_09,fernandez_10})
has a similar beneficial effect. Both instabilities help to keep the
accreted material in the gain region for a longer time before it is
advected deeper into the cooling region and onto the neutron star
surface \citep{buras_06_b,murphy_08b}.  If the advection time-scale
$\tau_\mathrm{adv}$ through the gain region (sometimes also termed
``residence time'' of the matter in the gain region) is increased
sufficiently and becomes comparable to or larger than the heating
time-scale $\tau_\mathrm{heat}$ required to unbind the material
between gain radius and shock, a runaway situation occurs, in which
neutrino heating leads to shock expansion, which in turn lengthens the
residence time, thus again increasing the neutrino heating efficiency
\citep{janka_98,thompson_00,janka_01b,buras_06_b,thompson_05,murphy_08b}.

However, whether an explosion can actually be brought underway in this
fashion in the most sophisticated supernova models with detailed
neutrino transport has not yet been finally and unambiguously
established on the basis of state-of-the-art neutrino hydrodynamics
simulations in 2D. Using their \textsc{Vertex-Prometheus} code, which
employs a variable Eddington factor technique to solve the neutrino
moment equations and the ``ray-by-ray-plus'' approach to cope with
multi-dimensional transport, the Garching group has found explosions
by the SASI-aided neutrino-driven mechanism for an $11.2 M_\odot$
solar mass progenitor \citep{buras_06_b,marek_09}, which can be
reproduced robustly for stiffer and softer nuclear equations of state,
(Marek~et~al. in preparation) and for a $15 M_\odot$ progenitor with
artificially imposed rotation at a rather late time $\approx 550
\ \mathrm{ms}$ after bounce. In other cases, i.e.\ for the same $15
M_\odot$ progenitor without rotation and with a better effective
gravitational potential, and for more massive progenitors, no
explosion was observed until $400 - 500 \ \mathrm{ms}$ after
bounce. In contrast to this, the Oak Ridge group, relying on a
multi-group flux-limited diffusion (MGFLD) algorithm combined with the
ray-by-ray-plus approach, has obtained explosions for a host of
different progenitors \citep{bruenn_06,bruenn_09}, while the former
Arizona group did not obtain neutrino-driven explosions at all with
their 2D MGFLD code \textsc{Vulcan} \citep{livne_04,livne_07} but
found acoustic explosions powered by sound waves created by violent
dipolar oscillations of the proto-neutron star
\citep{burrows_06,burrows_07}. Using yet another neutrino transport
scheme, the ``isotropic diffusion source approximation''
\citep{liebendoerfer_09} in the ray-by-ray-approximation and without
$\nu_\mu$ and $\nu_\tau$ transport, an explosion has also been
reported by \citet{suwa_10} and \citet{takiwaki_12} for an $11.2
M_\odot$ and a $13 M_\odot$ progenitor.

Given the disparity of methods and input physics, these
different results should not be overly surprising: None of the
aforementioned groups follow approaches that are completely identical
with respect to the treatment of neutrino transport (variable
Eddington factor method vs. MGFLD vs. IDSA, inclusion/omission of
non-isoenergetic scattering, velocity effects, and gravitational
redshift, ray-by-ray transport vs. multi-angle transport), the
neutrino processes (e.g.\ omission of $\mu/\tau$ neutrinos in
\citealt{suwa_10} and \citealt{takiwaki_12}), the hydro solver
(high-resolution shock capturing schemes vs. artificial viscosity), or
the treatment of gravity (effective relativistic potential
vs. Newtonian approximation) and nuclear burning (network vs. flashing
vs. omission of burning). A clear sensitivity of supernova
  simulations to Newtonian vs. GR gravity, the sophistication of the
  neutrino opacities, and observer corrections in the transport
  equations has recently been demonstrated in 1D models by
  \citet{lentz_12a}. Different results depending on the input physics
and the approximations used by the different groups are all the more
to be expected considering that some of the 2D models, whether
exploding or non-exploding, appear to be marginal cases (see
e.g.\ \citealp{marek_09}) anyway. From this perspective, the viability
of the neutrino-driven mechanism in 2D remains a controversial issue
and should still be investigated further with the help of better
simulations.

Considering the status of current 2D core-collapse supernova models,
it is conceivable that they still miss a crucial ingredient for robust
explosions. Alternatives to the standard neutrino-driven mechanism
have therefore been proposed and explored, such as
magnetohydrodynamically-driven explosions (see \citealt{dessart_07_a}
and references therein), energy input by acoustic waves
\citep{burrows_06,burrows_07}, or a phase transition to quark matter
\citep{sagert_09}. Moreover, 3D effects have recently been advocated
as the decisive factor by \citet{nordhaus_10} in their comparison of
1D, 2D, and 3D simulations using a strongly simplified prescription
for neutrino heating and cooling. Adopting the conceptual view of
\citet{burrows_93} of the neutrino-driven mechanism as a critical
phenomenon, they report a reduction of the so-called critical
luminosity in 3D by 15\% -- 25\% compared to 2D, and single out the
dimensionality as the key to successful supernova explosions,
discounting other factors such as general relativity and detailed
neutrino microphysics as minuscule corrections. However, whether these
results, obtained by means of a very rough approximation for neutrino
heating and cooling, can be verified by simulations with an elaborate
transport treatment is yet unclear (see
\citealp{hanke_11,takiwaki_12}). Moreover, the lower estimate of
\citet{nordhaus_10} for the importance of 3D effects relative to
  2D would probably not make them the single most important factor in
supernova physics, at least not by far. While there is no doubt that
3D models are indispensable for better understanding the SASI
\citep{blondin_07,iwakami_08,iwakami_09,fernandez_10}, the morphology
of supernova explosions, and the kicks and spins of neutron stars
\citep{hammer_10,wongwathanarat_10,rantsiou_10,fernandez_10}, our
understanding of supernovae certainly does not hinge on dimensionality
alone.

\emph{General relativity} (GR) is undoubtedly another major factor in
supernova physics due to the compactness of the neutron star ($M/R
\approx 0.1 \ldots 0.2$ in geometrized units) and the occurrence of
large infall ($v/c \approx -0.15 \ldots -0.3$) and outflow velocities.
In spherical symmetry (1D), the gauge freedom can be exploited to take
GR effects into account without sacrificing accuracy in the neutrino
transport sector more readily than in 2D and 3D; gray and multi-group
flux-limited diffusion schemes in 1D \citep{baron_89,bruenn_01} have
therefore been available since the 1980s and were followed by general
relativistic Boltzmann solvers a few years ago
\citep{yamada_99,liebendoerfer_01,liebendoerfer_04}. Comparisons with
the Newtonian case in 1D \citep{bruenn_01} clearly showed the
importance of GR effects, particularly at late times during the
accretion phase, where \citet{bruenn_01} find a much stronger
recession of the shock ($50\%$ smaller radius), higher luminosities,
in particular for electron neutrinos and antineutrinos (by $\approx
40\%$) and higher mean neutrino energies (by $\approx 15\%$) in
the GR case. This strong sensitivity to GR effects was
recently confirmed with up-to-date neutrino microphysics
by \citet{lentz_12a}. Considering the magnitude of relativistic effects, it is
obvious that they ought to be properly considered in multi-dimensional
supernova models as well.

Unfortunately, until recently the only
viable option to include GR effects in multi-dimensional neutrino
hydrodynamics simulations while retaining the framework of Newtonian
hydrodynamics has been the use of modified gravitational or
``effective'' potentials based on the Tolman-Oppenheimer-Volkov
equation of stellar structure
\citep{rampp_02,marek_06,mueller_08}. This ``pseudo-Newtonian''
approach is easy to implement, yields very satisfactory results in 1D
\citep{liebendoerfer_05,marek_06,mueller_10}, and certainly provides a
rough first-order approximation for GR effects in multi-dimensional
supernova models. As a complementary approach starting from
multi-dimensional general relativistic hydrodynamics simulations,
there have been efforts to address certain aspects of core-collapse
supernovae with the help of heavily simplified neutrino treatments
such as a ``deleptonization scheme'' \citep{liebendoerfer_05_b} or a
trapping scheme, but the applicability of such methods is naturally
limited, e.g.\ to the collapse and bounce phase,
\citep{ott_06_a,dimmelmeier_07_a,dimmelmeier_08}, or fast black hole
formation \citep{ott_11}.

In this paper, we pursue a considerably more ambitious course and
present the first 2D core-collapse supernova models combining general
relativistic hydrodynamics and a sophisticated energy-dependent
neutrino transport scheme. Using the \textsc{Vertex-CoCoNuT} code with
a ray-by-ray-plus treatment of multi-flavor, multi-energy 2D neutrino
transport as documented in \citet[henceforth Paper I]{mueller_10}, we
have conducted simulations of two progenitor models with $11.2 M_\odot$
and $15 M_\odot$ beyond the onset of the explosion. These simulations
are complemented by three non-exploding runs of the $15
M_\odot$ progenitor, viz. one with a pseudo-Newtonian (effective
potential) treatment of gravity, one with purely Newtonian gravity,
and one with a simplified set of neutrino interaction rates. The major
purpose of these simulations can be summarized as follows,
\begin{enumerate}
\item to demonstrate the feasibility of full-scale multi-dimensional
  general relativistic supernova simulations,
\item to evaluate the role of general relativistic effects in the
  explosion mechanism and the quality of the pseudo-Newtonian approach,
\item to gauge the sensitivity of the heating conditions to the
  neutrino physics input,
\item to closely reproduce and verify pseudo-Newtonian simulations
  with the \textsc{Prometheus-Vertex} code \citep{buras_06_b,marek_09}
  using a different hydrodynamics solver, thus eliminating
  reservations about the reliability of the numerics in existing 2D
  supernova models.
\end{enumerate}
Naturally, the impact of GR on the gravitational wave (and also on the
neutrino) signal from core-collapse supernovae is also a question of
paramount importance.  However, this issue cannot be discussed within
the scope of the present paper with its focus on the \emph{dynamics}
of the post-bounce evolution, and will be the subject of a follow-up
publication.

To address the aforementioned issues, we have structured our paper in
the following manner: We review the numerical scheme and outline the
model setup and input physics used in this work in
Sections~\ref{sec:numerics} and \ref{sec:setup}. In
Section~\ref{sec:results_dynamics}, we describe the shock evolution and
the explosion morphology of our relativistic explosion models.  In
Section~\ref{sec:heating}, we then address differences to the
non-exploding runs and provide a more quantitative analysis of the
heating conditions for our models in order to determine the impact of
the GR treatment and the neutrino microphysics on the post-bounce
dynamics. We summarize the main results from this analysis in
Section~\ref{sec:summary} and evaluate the implications with respect to
the major issues of our paper.

\section{Numerical Methods}
\label{sec:numerics}
We perform numerical simulations with the general relativistic
neutrino hydrodynamics code \textsc{Vertex-CoCoNuT} introduced in
Paper~I \citep{mueller_10}, which is a combination of the neutrino
transport solver \textsc{Vertex} \citep{rampp_02,buras_06_a} and the
relativistic hydrodynamics code \textsc{CoCoNuT}
\citep{dimmelmeier_02_a,dimmelmeier_04}. \textsc{CoCoNuT} is a
time-explicit, directionally unsplit Eulerian Godunov-type
finite-volume solver written for spherical polar coordinates and uses
piecewise parabolic (PPM) reconstruction and Runge-Kutta time-stepping to achieve
higher-order spatial and temporal accuracy. Our implementation of
\textsc{CoCoNuT} employs a relativistic version of the HLLC
approximative Riemann solver \citep{mignone_05_a}, but adaptively
switches to the more diffusive HLLE solver \citep{einfeldt_88} in the
vicinity of strong shocks to avoid the phenomenon of
odd-even-decoupling \citep{quirk_94}. In order to reduce spurious
numerical energy generation, we use an improved formulation of the
energy equation described in Paper~I. The metric equations are solved
approximatively using the extended conformal flatness condition (xCFC,
\citealp{cordero_09}), a constrained scheme with improved numerical
stability properties compared to the original conformally flat
approximation of \citet{isenberg_78}. In the context of core-collapse,
this approximation works extremely well as demonstrated by the
excellent agreement between rotational core-collapse simulations using
the CFC approximation and those relying on the full ADM formalism
\citep{ott_06_a,ott_06_b,dimmelmeier_07_b}.

The time-implicit neutrino transport module \textsc{Vertex} integrates
the moment equations for the neutrino energy and momentum density
using a variable Eddington factor technique \citep{rampp_02}. The
higher moments of the neutrino radiation field that are required to
close the system of moment equations are obtained from a
simplified Boltzmann equation that is solved in conjunction with the
neutrino moment equations within a fixed-point iteration. All
velocity- and metric-dependent terms are fully included in the moment
equations, as is energy redistribution by non-isoenergetic
scattering. In 2D, we resort to the ``ray-by-ray-plus'' approximation
\citep{buras_06_a,bruenn_06}, assuming that the neutrino distribution
function is axially symmetric around the radial direction (which
implies a radial flux vector). This allows us to solve independent 1D
transport problems along ``rays'' corresponding to the angular zones
of the polar grid. However, the lateral advection of neutrinos and the
effect of lateral neutrino pressure gradients are both included to
avoid unphysical behavior in the optically thick regime
\citep{buras_06_a}.

We use an up-to-date set of neutrino interaction rates in
\textsc{Vertex}; in addition to the ``standard'' set of neutrino
opacities from \citet{bruenn_85}, we take nucleon-nucleon
bremsstrahlung \citep{hannestad_98}, neutrino-neutrino pair conversion
\citep{buras_03} and inelastic neutrino scattering off heavy nuclei
\citep{langanke_08} into account. Furthermore, we compute electron
captures on heavy nuclei using the improved rate tables of
\cite{langanke_03} instead of the FFN rates \citep{fuller_82}, and as
an alternative to the `isoenergetic'' (i.e.\ no energy exchange with
nucleons treated as vacuum particles) approximation of
\citet{bruenn_85} we include recoil effects, high-density correlations
\citep{burrows_98,burrows_99}, and weak magnetism corrections
\citep{horowitz_97} in our treatment of charged and neutral-current
neutrino-nucleon interactions (see Table~\ref{tab:rates}).

\section{Model Setup}
\label{sec:setup}
\subsection{Progenitors and Neutrino Physics}
In this paper, we consider two different non-rotating progenitors,
namely the models s11.2 of \citet{woosley_02} and s15s7b2 of
\citet{woosley_95}. These $11.2 M_\odot$ and $15 M_\odot$ progenitors
have been chosen such as to facilitate a comparison with the
pseudo-Newtonian simulations of \citet{buras_06_b} and
\citet{marek_09}. The $15 M_\odot$ case is of particular interest
since \citet{marek_09} observed an explosion for this progenitor in a
run with artificially imposed rotation and an overly strong effective
gravitational potential (their model LS-rot) while the corresponding
non-rotating model (LS-2D), computed with the best effective potential
of \citet{marek_06}, failed to explode before the end of their
high-resolution run $420 \ \mathrm{ms}$ after bounce. Within the model
assumptions of \citet{marek_09}, the progenitor s15s7b2 is therefore a
marginal case, and hence ideally suited for investigating possible
effects of general relativity and testing the accuracy of the
effective potential approach. In addition to two relativistic models
(G11 and G15) for the two progenitors computed with
  \textsc{Vertex-CoCoNuT}, we therefore also consider a purely
Newtonian (N15) and a pseudo-Newtonian (M15) simulation of the $15
M_\odot$ star of \citet{woosley_95} computed with
  \textsc{Vertex-Prometheus}. 

Furthermore, we also include another
relativistic calculation (S15) with a slightly simplified set of
neutrino opacities to assess the importance of improved interactions
rates (particularly for neutrino-nucleon processes). A summary of the
input physics for these models is given in
Table~\ref{tab:model_setup}, and the differences between the full and
simplified set of neutrino rates are given in detail in
Table~\ref{tab:rates}.
Model S15 serves to illustrate the
possible variations in 2D core-collapse supernova simulations
that may be due the treatment of the neutrino microphysics.
Since a full investigation of all the \emph{individual} rates in
multi-dimensional supernova simulations is not feasible
for lack of computer power, we choose a ``package'' of opacities
for model S15 that is roughly representative for the
neutrino treatment used in the 1980s and 1990s and, together
with model G15, spans a reasonable part of the range of
sophistication of the neutrino microphysics introduced in modern
multi-dimensional core-collapse simulations. While model
S15 thus provides some rough indications about the
influence of the neutrino rates in core-collapse supernovae,
it should be borne in mind that even more radical approximations
than our ``simplified'' set of interaction rates are used
(e.g.\ $\mu$- and $\tau$- neutrinos are sometimes ignored completely,
or treated by a leakage/trapping scheme),
which can affect the dynamics much more seriously.

With the exception of model N15 (where 128 angular zones were used),
all runs were performed on a spherical polar grid covering $180^\circ$
in latitude with $400 \times 64$ zones initially. We simulate the
interior of each progenitor out to $10 000 \ \mathrm{km}$,
corresponding to a mass coordinate of $1.57 M_\odot$ (s11.2) and $2.0
M_\odot$ (s15s7b2), respectively. The distribution of the radial grid
varies between the individual simulations\footnote{Specifically,
  stability considerations require different zoning in the
  hydrodynamics modules \textsc{CoCoNuT} and \textsc{Prometheus} near
  the origin of the spherical polar grid.}, but was chosen (and when
necessary re-adjusted) such that the density gradient at the surface
of the proto-neutron star remains sufficiently well resolved during
the simulation. With a moderate angular resolution of $2.8^\circ$ we
have settled for an affordable compromise in the first long-time
multi-dimensional simulations using the relativistic
\textsc{Vertex-CoCoNuT} code. For future calculations (relying on more
efficient parallelization of the hydrodynamics and metric solvers), a
higher resolution is clearly desirable, but past experience with
pseudo-Newtonian simulations suggests that 64 angular zones already
provide a reasonable level of accuracy to study systematic differences
between models with different input physics. Both for the $11.2
M_\odot$ and the $15 M_\odot$ progenitor, this resolution is in fact
sufficient for good quantitative agreement with high-resolution
studies in the pseudo-Newtonian case as demonstrated by the similarity
of our model M15 with model LS-2D of \citealt{marek_09}. For
comparisons or extrapolation, it should nonetheless be borne in mind
that runs with higher angular resolution generally appear somewhat
more optimistic in 2D (cp.\ \citealp{scheck_phd} for numerical tests
with approximate, gray neutrino transport and \citealp{hanke_11} for
recent tests with simple neutrino source terms). The reason for this
empirical trend is yet to be determined, although \citet{hanke_11}
suggest that the inverse turbulent energy cascade in 2D, which may
shift energy from high-$\ell$ convective modes into low-$\ell$ SASI
modes, could provide an explanation. The effect could be
connected to a reduction of the dissipation on small scales
with higher resolution or to additional energy input into
convective motions on the smallest availables scales.
 We refer the reader to their
paper for a more thorough discussion of this issue.

For the neutrino transport, we chose a logarithmically-spaced energy
grid with 12 bins ranging from $0 \ \mathrm{MeV}$ to $380
\ \mathrm{MeV}$, which -- though not an optimal resolution -- yields
fairly similar dynamics compared to a finer zoning in energy space
(cp.\ \citealp{marek_09} for an example with only 9 bins).

\begin{table*}
  \caption{Model Setup
    \label{tab:model_setup}
  }
  \begin{center}
    \begin{tabular}{ccccccc}
      \hline \hline 
            &             & neutrino  & hydro  & treatment of       & final post-bounce       & angular  \\
      model & progenitor  & opacities & solver & relativity         & time reached\tablenotemark{a}    & resolution \\
      \hline
      G11 & s11.2   & full set    & \textsc{CoCoNuT} & GR hydro + xCFC    &  $920 \ \mathrm{ms}$  & $2.8^\circ$ \\
      G15 & s15s7b2 & full set    & \textsc{CoCoNuT} & GR hydro + xCFC    &  $775 \ \mathrm{ms}$  & $2.8^\circ$ \\ 
      S15 & s15s7b2 & reduced set & \textsc{CoCoNuT} & GR hydro + xCFC    &  $474 \ \mathrm{ms}$  & $2.8^\circ$ \\
      M15 & s15s7b2 & full set    & \textsc{Prometheus} & Newtonian +
                                    modified potential\tablenotemark{b}  &  $517 \ \mathrm{ms}$  & $2.8^\circ$ \\ 
      N15 & s15s7b2 & full set    & \textsc{Prometheus} & Newtonian (purely) &  $525 \ \mathrm{ms}$  & $1.4^\circ$ \\ 
      \hline \hline
    \end{tabular}
    \tablenotetext{1}{In practice, technical reasons limit the
      simulation time. The five simulations discussed here all needed
      to be terminated because convergence of the implicit transport
      solver could only be ensured by inordinately small time-steps at
      late times. Such convergence problems occur because of extremely
      steep velocity gradients in spots where fast downflows strike
      the proto-neutron star surface.} 
      \tablenotetext{2}{Case~A of \citet{marek_06}}
  \end{center}
\end{table*}

\begin{table*}
  \caption{Neutrino physics input
    \label{tab:rates}
  }
  \begin{center}
    \begin{tabular}{ccc}
      \hline \hline & full rates       & simplified rates \\
           process  & (G11, G15, M15, N15)  &       (S15) \\
      \hline
      $\nu A \rightleftharpoons \nu A$ & \citet{horowitz_97} (ion-ion correlations) & \citet{bruenn_97} \\
                                       & \citet{langanke_08} (inelastic contribution)& \citet{itoh_04} (ion-ion-correlations)\\
      $\nu \ e^\pm \rightleftharpoons \nu \ e^\pm$  &  \citet{mezzacappa_93} & \citet{mezzacappa_93}\\
      $\nu \ N \rightleftharpoons \nu \ N$          & \citet{burrows_98}\tablenotemark{a}&  \citet{bruenn_85}\\
      $\nu_e\ n \rightleftharpoons e^-\ p$      & \citet{burrows_98}\tablenotemark{a}&  \citet{bruenn_85}\\
      $\bar{\nu}_e \ p \rightleftharpoons e^+ \ n$  & \citet{burrows_98}\tablenotemark{a}&  \citet{bruenn_85}\\        
      $\nu_e \ A' \rightleftharpoons e^- \ A$  & \citet{langanke_03}&  \citet{fuller_82,bruenn_85}\\      
      $\nu\bar{\nu} \rightleftharpoons e^- \ e^+ $ & \citet{bruenn_85,pons_98} & \citet{bruenn_85,pons_98} \\
      $\nu\bar{\nu} \ NN \rightleftharpoons NN $ & \citet{hannestad_98} & \citet{hannestad_98} \\
      $\nu_{\mu,\tau}\bar{\nu}_{\mu,\tau} \rightleftharpoons \nu_{e}\bar{\nu}_{e} $ 
                                                    & \citet{buras_03} & --- \\
$\stackrel{(-)}{\nu}_{\mu,\tau}\stackrel{(-)}{\nu}_e \rightleftharpoons \stackrel{(-)}{\nu}_{\mu,\tau}\stackrel{(-)}{\nu}_e$ 
                                                    & \citet{buras_03} & --- \\                  
      \hline \hline 
    \end{tabular}
    \tablenotetext{1}{ Note that these reaction rates account for
      nucleon thermal motions, phase-space blocking, energy transfer
      to the nucleon associated with recoil (``non-isoenergetic''
      scattering), and nucleon correlations at high densities.
      Moreover, we include the quenching of the axial-vector coupling
      at high densities \citep{carter_02}, correction to the effective
      nucleon mass \citep{reddy_99}, and weak magnetism effects
      \citep{horowitz_02} in our full set of rates.}
  \end{center}
\end{table*}

\subsection{Equation of State}
\begin{figure*}
\plottwo{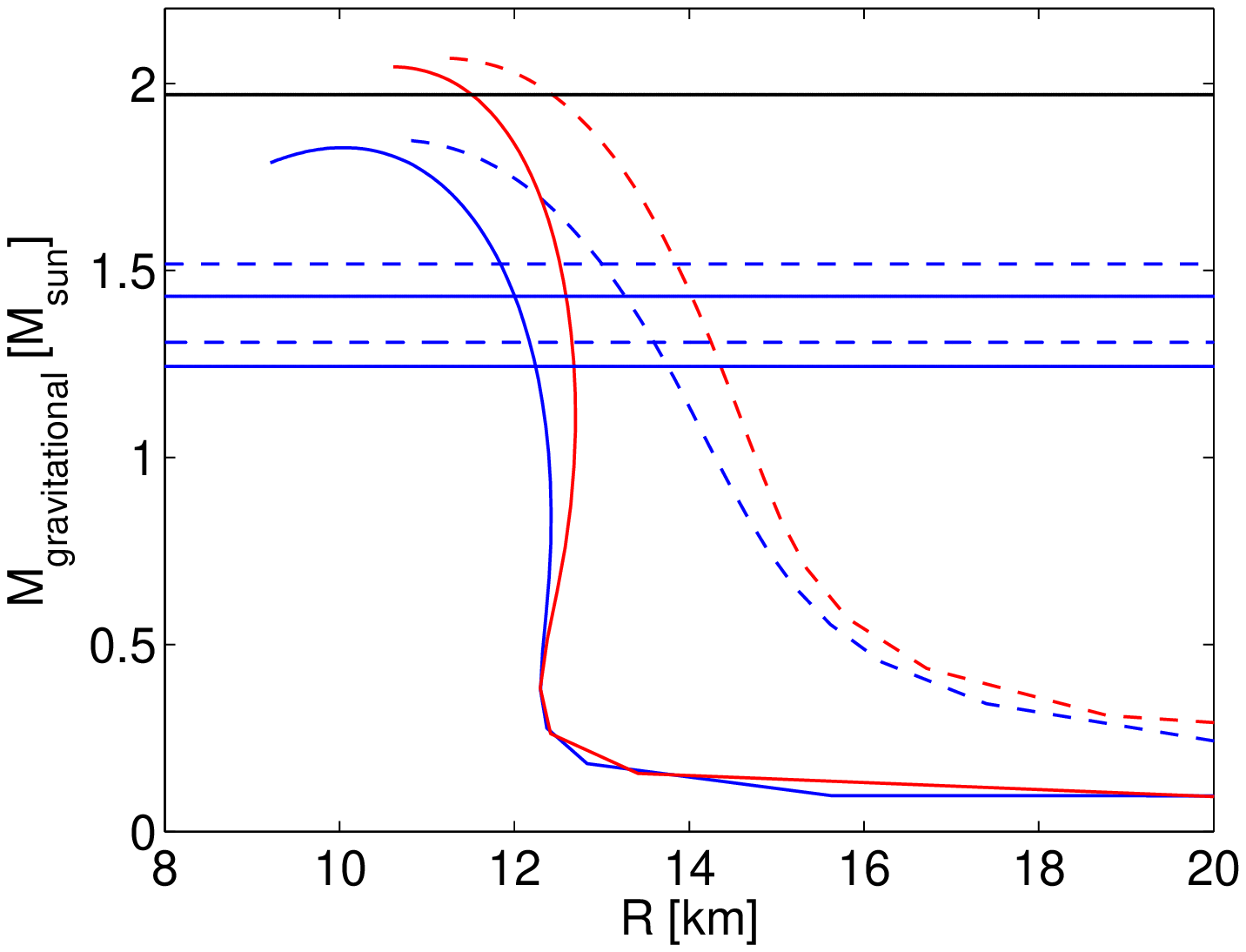}{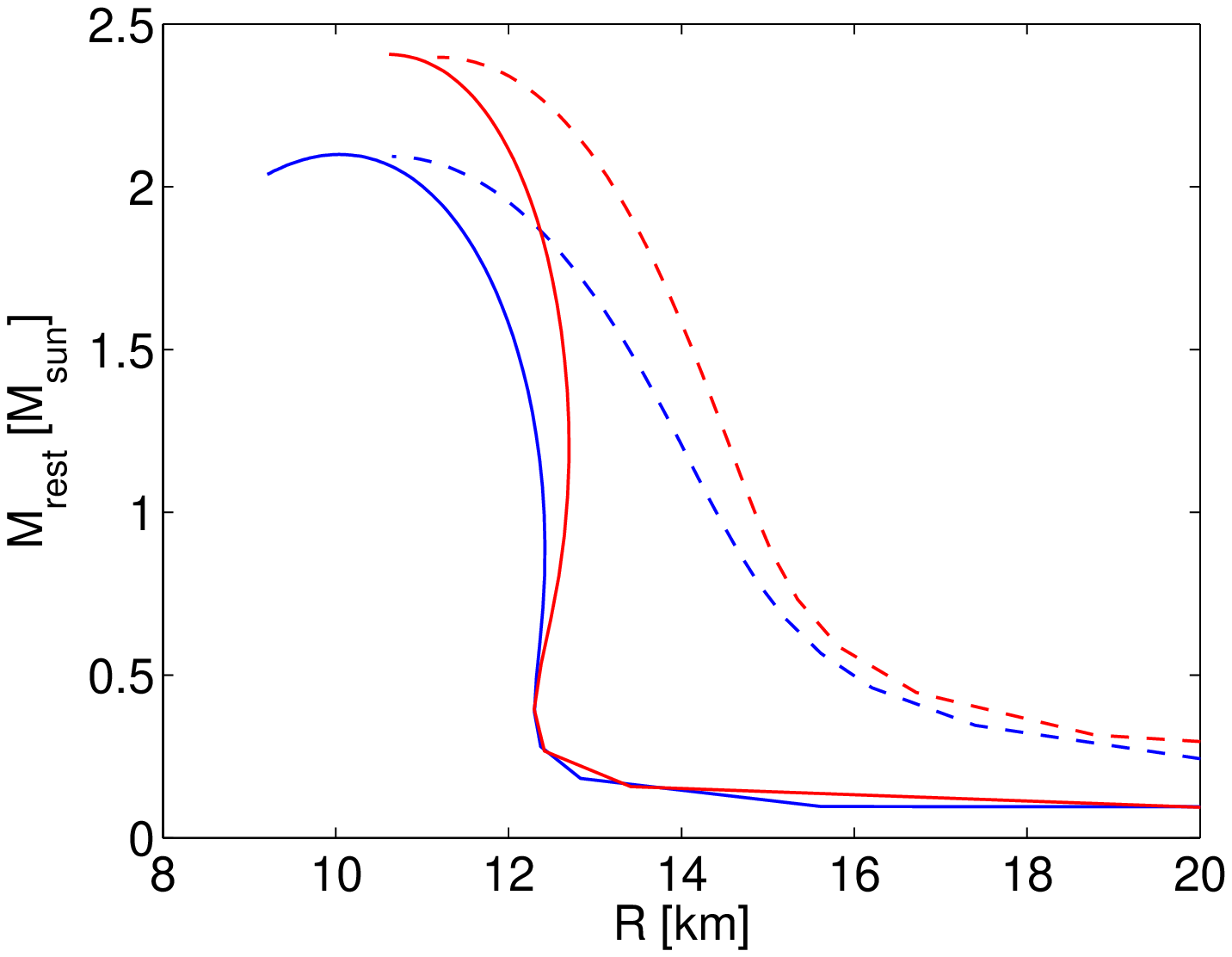}
\caption{Mass-radius relations of the equations of state LS180 (blue)
  and LS220 (red) for the gravitational mass (left panel) and the
  baryonic mass (right panel). Solid lines display the case of cold
  neutron stars ($T=0$), while curves for the case of a hot
  proto-neutron star with a constant entropy of $s=1.5
  \ \mathrm{k}_b/\mathrm{nucleon}$ are shown as dashed lines. The
  black horizontal line in the left panel corresponds to a mass of
  $1.97 M_\odot$ as measured by \citet{demorest_10} for the pulsar
  J1614-2230. The gravitational masses for neutron stars with baryonic
  masses of $1.36 M_\odot$ and $1.58 M_\odot$ are indicated both for
  $T=0$ (solid blue horizontal lines) or $s=1.5
  \ \mathrm{k}_b/\mathrm{nucleon}$ (dashed blue horizontal lines) in
  the left panel (figures provided by A.~Bauswein.)
\label{fig:ns_mr} }
\end{figure*}

The simulations were performed using a soft version of the equation of
state (EoS) of \citet{lattimer_91} with a bulk incompressibility
modulus of nuclear matter of $K=180 \ \mathrm{MeV}$ (LS180). The value
of $K$ in this EoS has raised some objections
\citep{nordhaus_10,dasgupta_12} because measurements point to $K=240
\ \mathrm{MeV}$ for symmetric nuclear matter \citep{shlomo_06}, and
because observations have recently pushed the maximum neutron star
mass to at least $2 M_\odot$ \citep{demorest_10}, which is
incompatible with this particular EoS (see Figure~\ref{fig:ns_mr}
and cp.\ also \citealp{hempel_12}).

However, this criticism neglects that (at least for the range of
progenitors we are considering here) the crucial quantity for the
development of the explosion is not the maximum neutron star mass, but
rather the radius of the proto-neutron star during the reheating
phase.  For the baryonic neutron star masses we obtain at the end of
our longest-running simulations ($1.35 M_\odot$ and $1.58 M_\odot$ for
the $11.2 M_\odot$ and $15 M_\odot$ progenitor, respectively), the
LS180 EoS yields a final radius (in the cold state) of around $12
\ \mathrm{km}$ (see Figure~\ref{fig:ns_mr}), which is smaller by only
$0.5 \ldots 0.7 \ \mathrm{km}$ than the neutron star radius predicted
by the ("stiffer") LS220 EoS (which has $K = 220 \ \mathrm{MeV}$ and
yields a maximum neutron star mass larger than $2 M_\odot$).
Figure~\ref{fig:ns_mr} also shows that the deviation is of similar
magnitude for hot proto-neutron stars.  Moreover, a neutron star
radius of $12 \ \mathrm{km}$ is well compatible with observed neutron
star radii; it actually agrees nicely with the best existing
(observational and theoretical) estimates of neutron star radii, see
e.g.\ the papers by \citet{steiner_10} and \citet{hebeler_10}, a fact
that, for example, disfavors the EoS of \citet{shen_98}, because the
latter yields a radius of $\approx 15 \ \mathrm{km}$ for a neutron
star with a gravitational mass of $1.4 M_\odot$. The prospective
differences to a simulation with LS220 become even smaller if we
consider that the two neutron stars are even less massive when the
explosion develops (e.g.\ at a baryonic mass of $1.51 M_\odot$ for the
$15 M_\odot$ progenitor).

It is also important to note that the neutron stars formed in the
$11.2 M_\odot$ and $15 M_\odot$ runs have \emph{baryonic} masses of $1.35
M_\odot$ and $1.58 M_\odot$, but the corresponding gravitational
masses are about $\approx 10 \%$ lower than these values. These masses
are far below the maximum gravitational mass that can be supported by
the LS180 EOS (which is about $1.85 M_\odot$, corresponding to a
baryonic mass well beyond $2 M_\odot$; Figure~\ref{fig:ns_mr}).

In the neutron star mass regime we are dealing with in this paper,
LS180 therefore remains a justifiable choice for the EoS. This
``softer EoS'' yields neutron star radii quite similar to the
``stiffer'' version LS220 because of relatively small differences in
the pressure-density relation.  It is also very similar in most other
EoS properties. Sizable differences appear only close to the limiting
neutron star masses supported by these equations of state.

\begin{figure}
\plotone{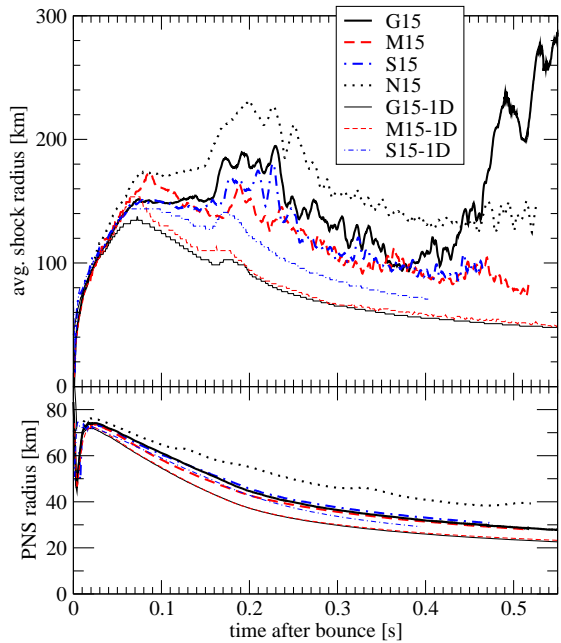}
\caption{Average shock radius and proto-neutron star (PNS) radius
  (defined by a fiducial density of $10^{11} \ \mathrm{g}
  \ \mathrm{cm}^{-3}$) for the 2D models G15 (GR, full rates, black
  thick solid line), S15 (GR, reduced rates, blue, thick,
  dash-dotted), M15 (pseudo-Newtonian, full rates, red, thick,
  dashed), and M15 (purely Newtonian, black, thick, dotted). 1D models
  corresponding to G15, M15, and S15 are also shown as thin lines for
  comparison. Note that the shock is located considerably further out
  in S15-1D than in G15-1D and M15-1D. This is a consequence of the
  strong sensitivity of the shock position $r_\mathrm{sh}$ to the PNS
  radius, $r_\mathrm{PNS}$, for a stationary spherical accretion flow
  ($r_\mathrm{sh} \propto r_\mathrm{PNS}^{8/3}$, see, e.g.,
  Equation~(1) of \citealt{marek_09}).  The larger PNS radius in
  S15-1D can in turn be traced to less efficient cooling by $\mu/\tau$
  neutrinos and higher temperatures in the density region
  $10^{12}\ldots 10^{13} \ \mathrm{g} \ \mathrm{cm}^{-3}$. Different
  PNS radii (caused by proto-neutron star convection; see Appendix~C
  in \citealt{buras_06_b}) are also responsible for the larger shock
  radii in the 2D models G15 and M15 compared to G15-1D and M15-1D at
  early times, when multi-dimensional effects in the gain region do
  not yet play a significant role. (The data for M15-1D have been
  provided by L.~H\"udepohl.)
\label{fig:s15_shock_radius}
}
\end{figure}

\section{Explosion Dynamics and Energetics}
\label{sec:results_dynamics}
The relativistic supernova simulations of the $11.2 M_\odot$ and $15
M_\odot$ stars discussed in this paper both yield successful
explosions.  In this section, we address the evolution of models G11
and G15 in a descriptive manner, focusing on the propagation of the
shock, the activity of the SASI, the morphology of the explosion,
and the ejecta composition.
The heating conditions in our models and the crucial differences
responsible for the different outcome of the $15 M_\odot$ models G15,
M15, N15, and S15 will be analyzed in the next section.

\label{sec:explosion}

\begin{figure*}
\plottwo{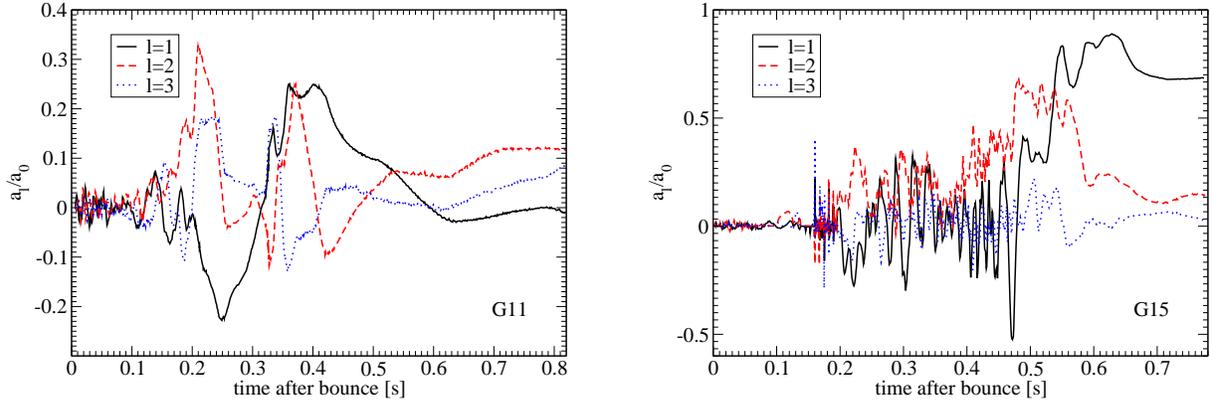}{f3b.eps}
\caption{First, second, and third coefficient $a_1$, $a_2$, and
  $a_3$, of the decomposition of the shock radius into Legendre
  polynomials, normalized to the coefficient $a_0$ (i.e.\ the average
  shock radius) for model G11 (left panel) and model G15 (right
  panel).
\label{fig:sasi}
}
\end{figure*}

\begin{figure*}
\plottwo{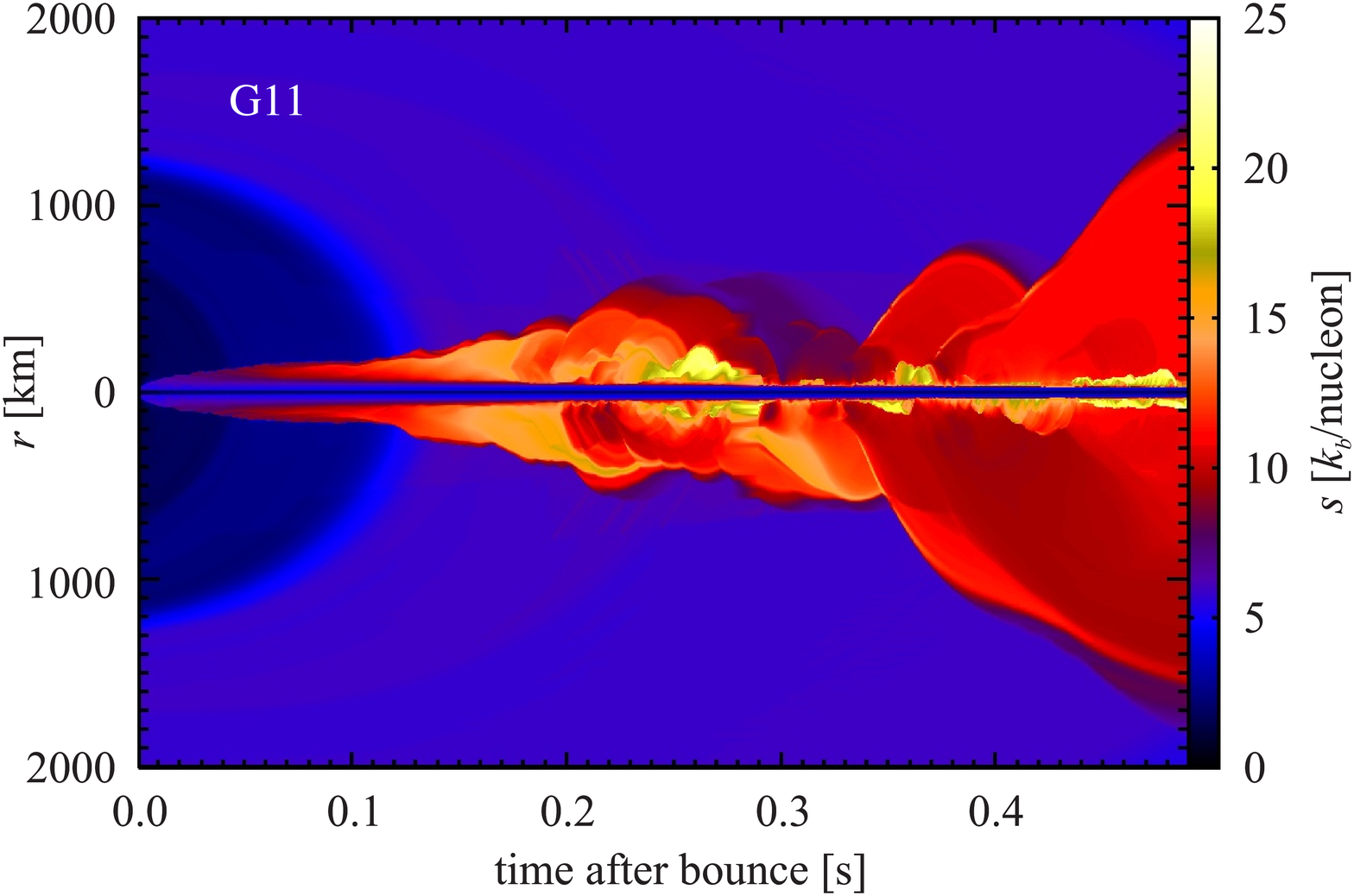}{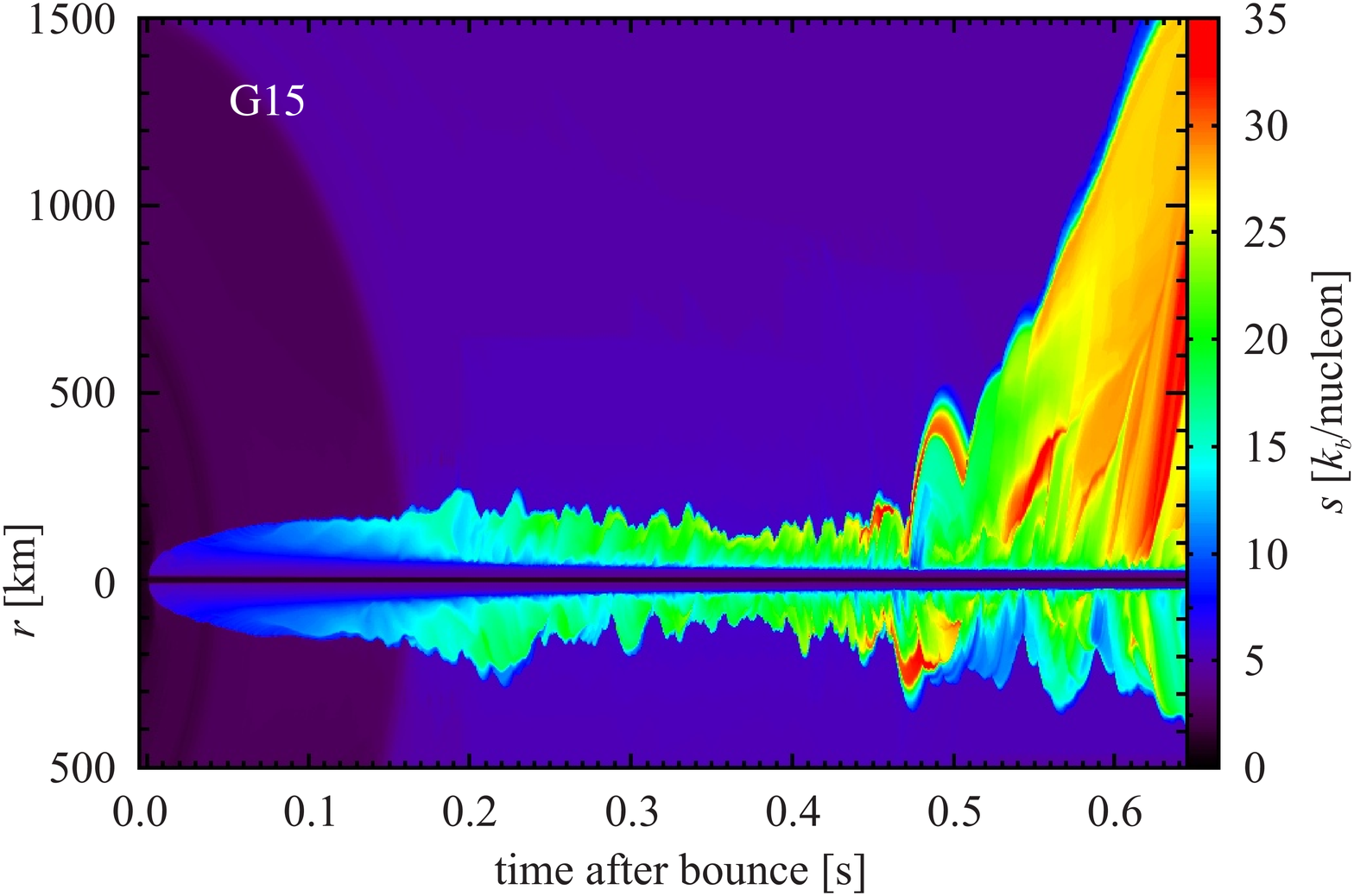}
\caption{Entropy along the north and south polar axis as a function of
  time for models G11 (left) and G15 (right). The trajectory of the
  Si/SiO-interface -- initially located at $\approx 1200
  \ \mathrm{km}$ and $\approx 1800 \ \mathrm{km}$, respectively -- is
  visible as a discontinuity in brightness. We point out that model
  G11 (left) still shows strong SASI activity with repeated phases of
  shock stagnation or retraction during its long approach to the
  explosion. It should also be noted that the entropy of the buoyant
  bubbles pushing the shock further out in the polar directions is
  quite moderate in this case. Model G15 (right) shows a strong
  dipolar asymmetry during the explosion phase: While neutrino-heated
  high-entropy plumes have pushed the shock outward with a propagation
  velocity of $\gtrsim 10000 \ \mathrm{km} \ \mathrm{s}^{-1}$ along
  the northern polar direction, the shock radius in the opposite
  direction grows much more slowly and non-monotonically. Note
  that only parts of the simulations are covered by the plots because
  many features of the earlier phases would not be recognizable on a
  different scale.
\label{fig:entropy_poles}
}
\end{figure*}

\begin{figure*}
\plottwo{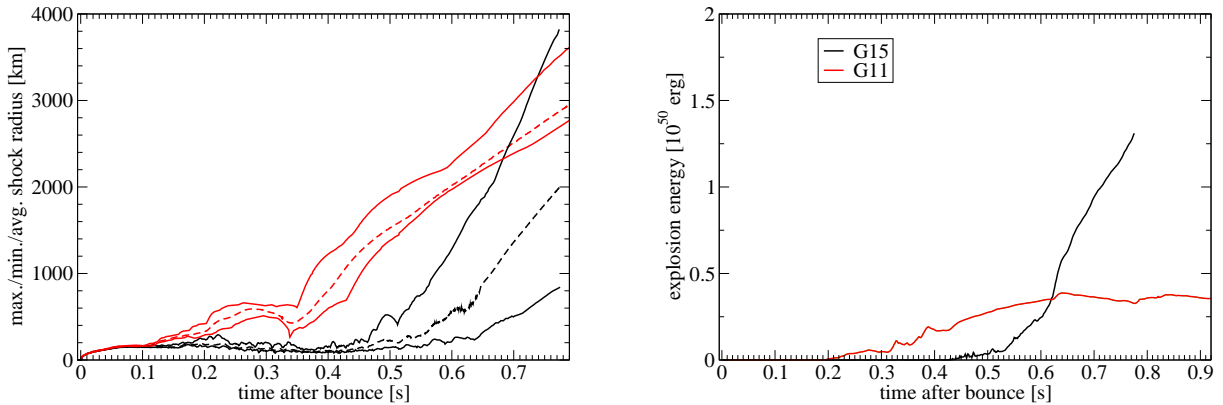}{f5b.eps}
\caption{Left panel: Maximum, minimum and average shock radius for
  models G11 and G15. In both cases, the shock expands and SASI
  activity increases considerably when the Si/SiO-interface reaches
  the shock (at $\approx 100 \ \mathrm{ms}$ and $\approx 150
  \ \mathrm{ms}$, respectively). It recedes again in the case of model
  G15, and the explosion is launched only several hundreds of
  $\mathrm{ms}$ later.  Right panel: The diagnostic ``explosion
  energy'' for both models (see text for exact definition).
\label{fig:shock_radius_and_explosion_energy}
}
\end{figure*}

\begin{figure*}
\plottwo{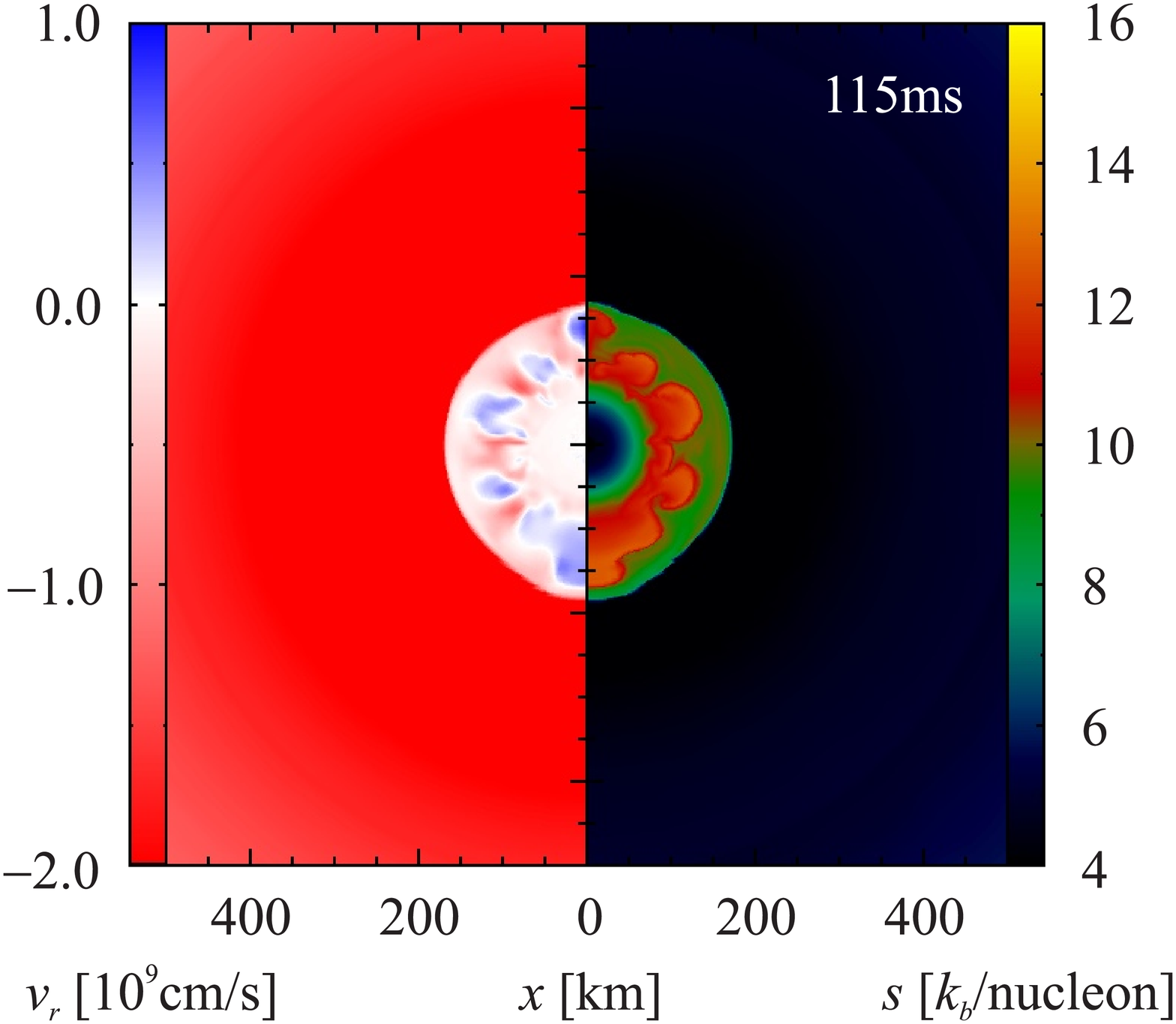}{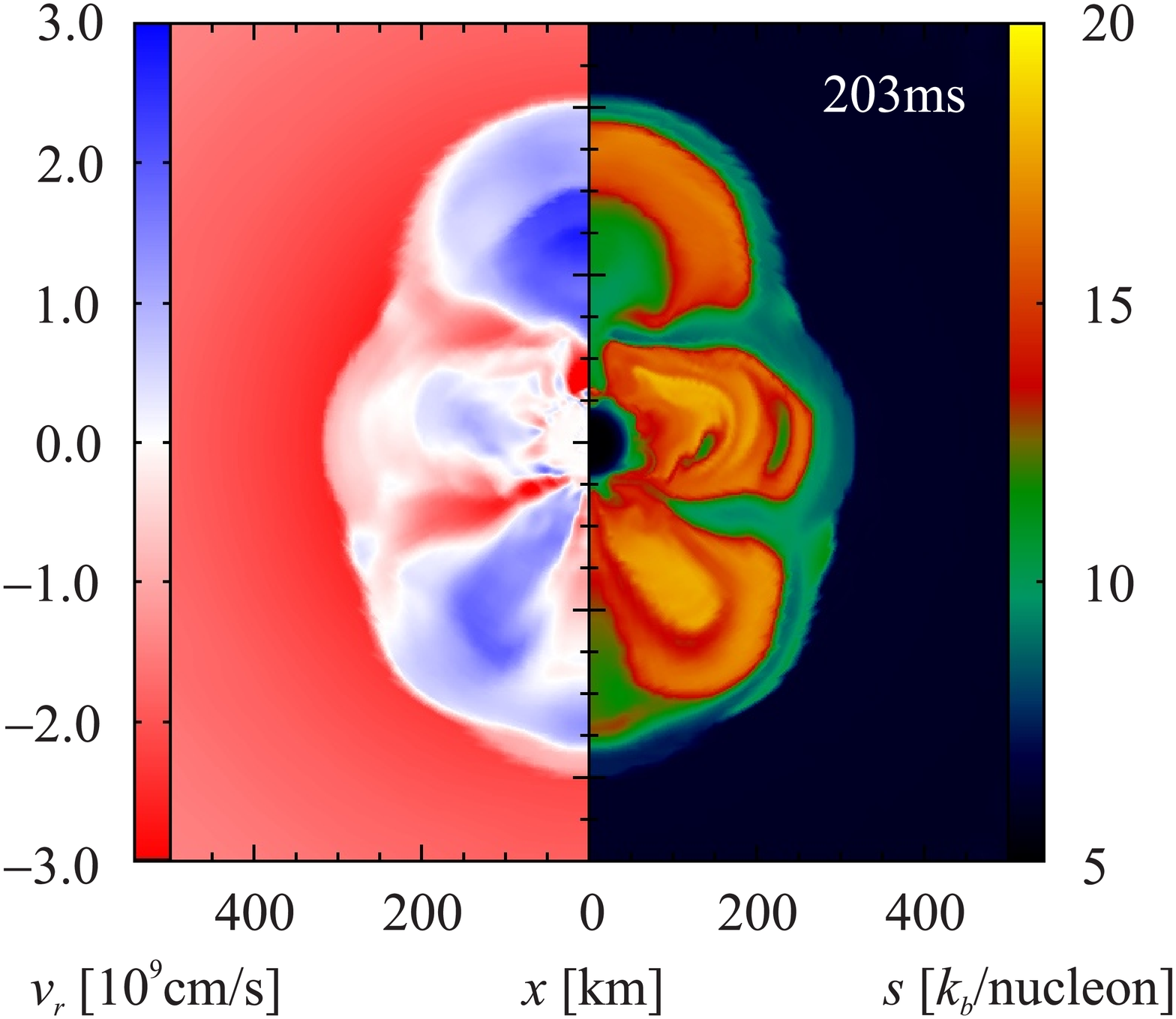}
\vspace{0.2in}
\plottwo{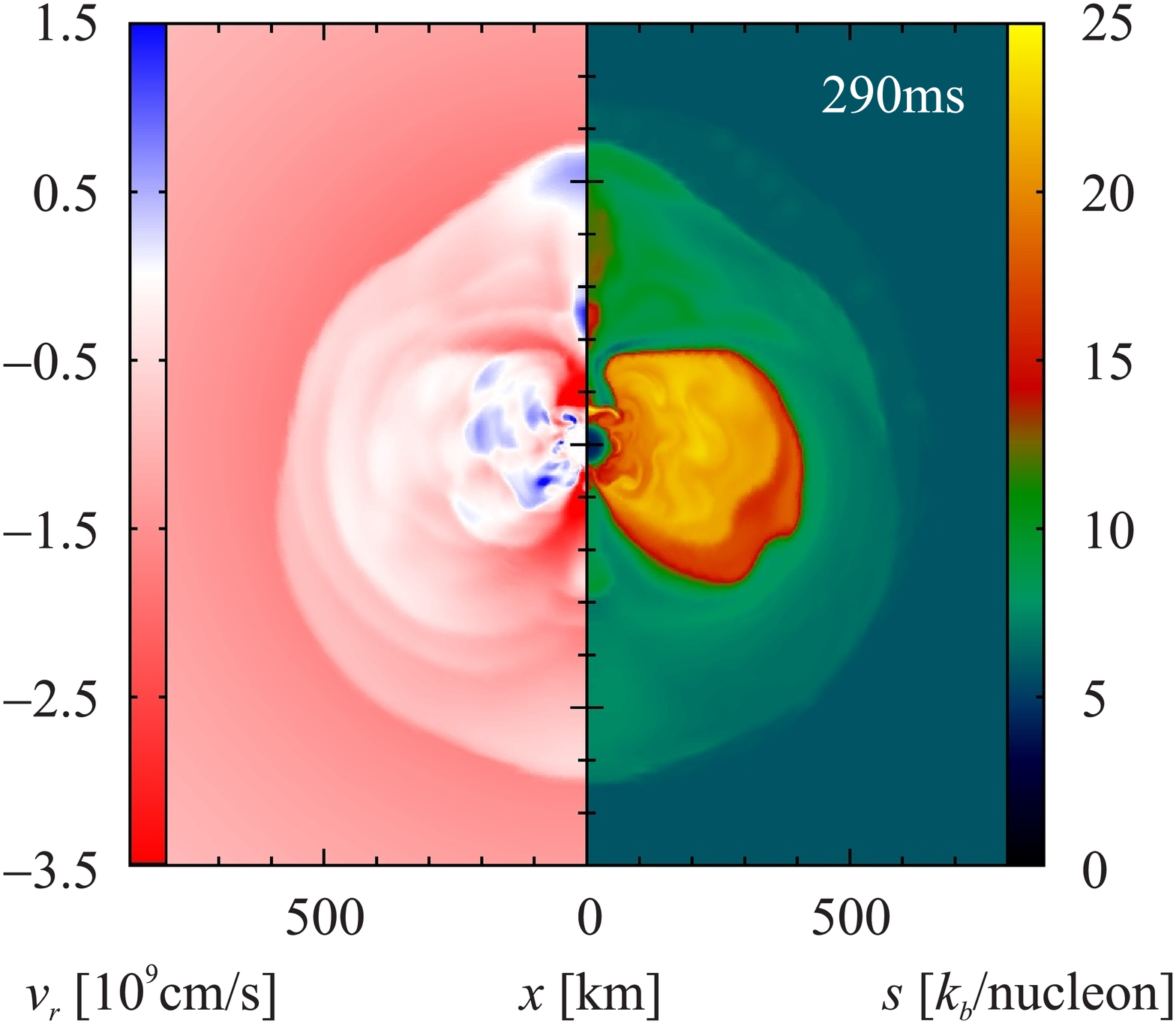}{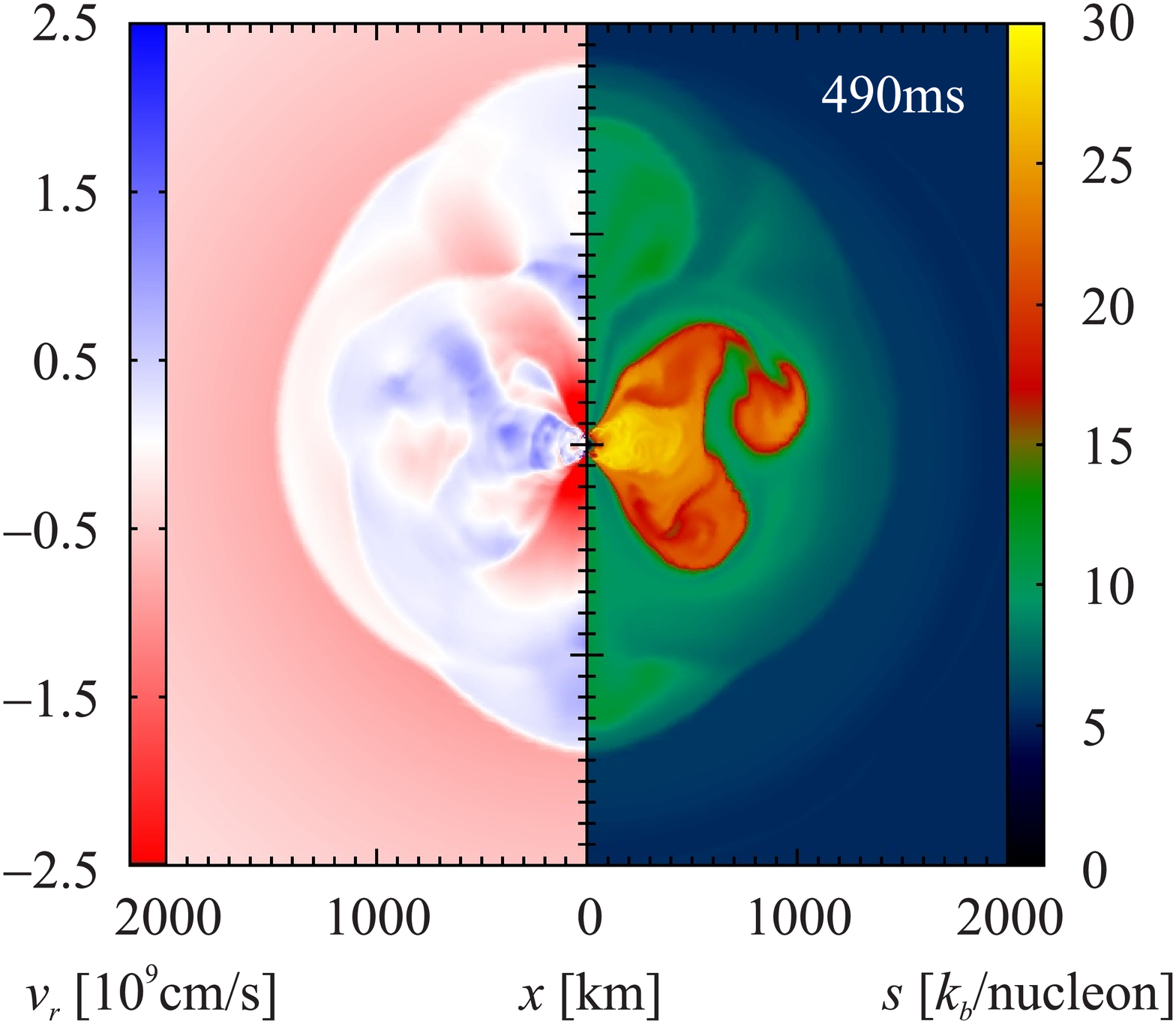}
\vspace{0.2in}
\plottwo{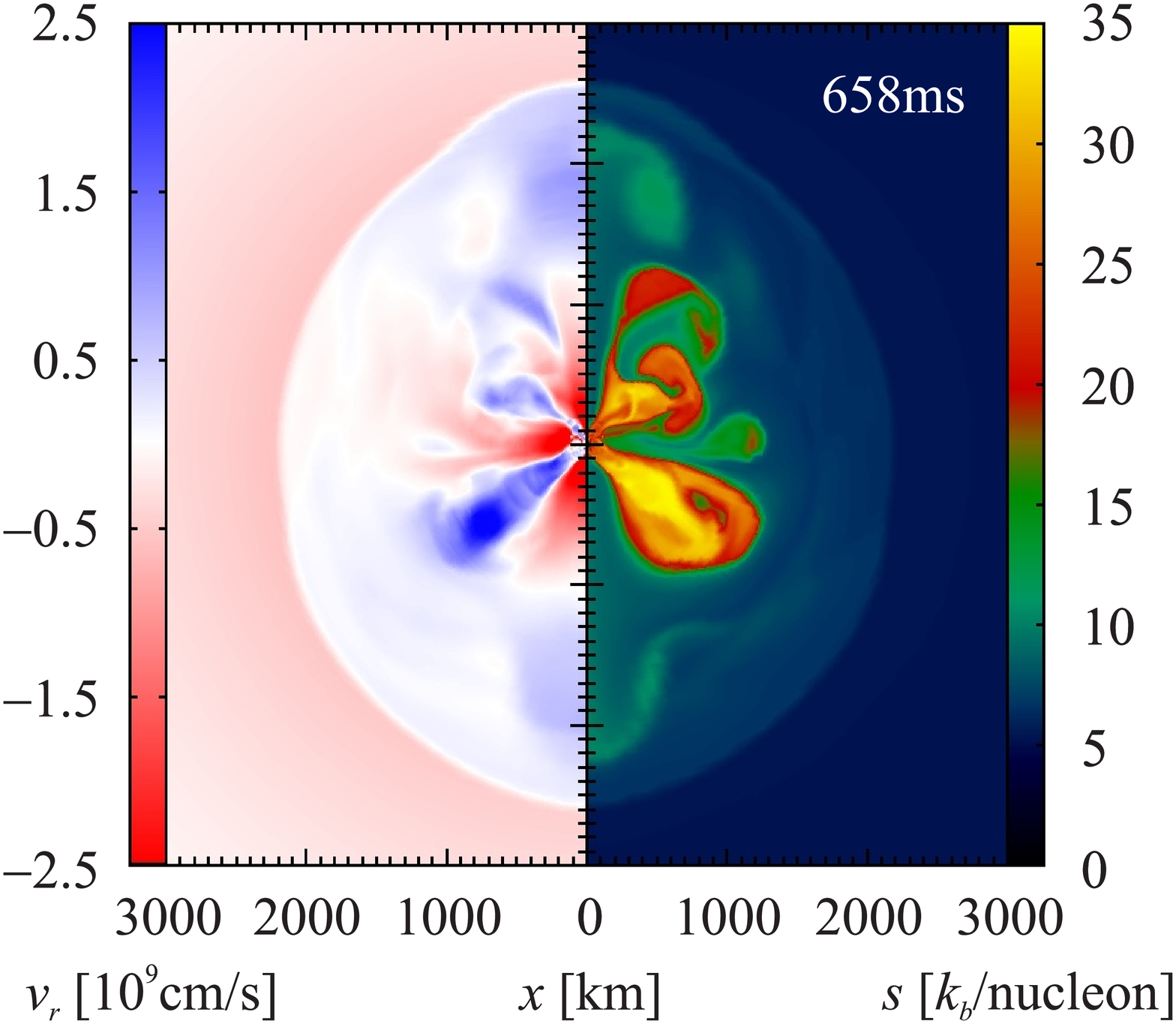}{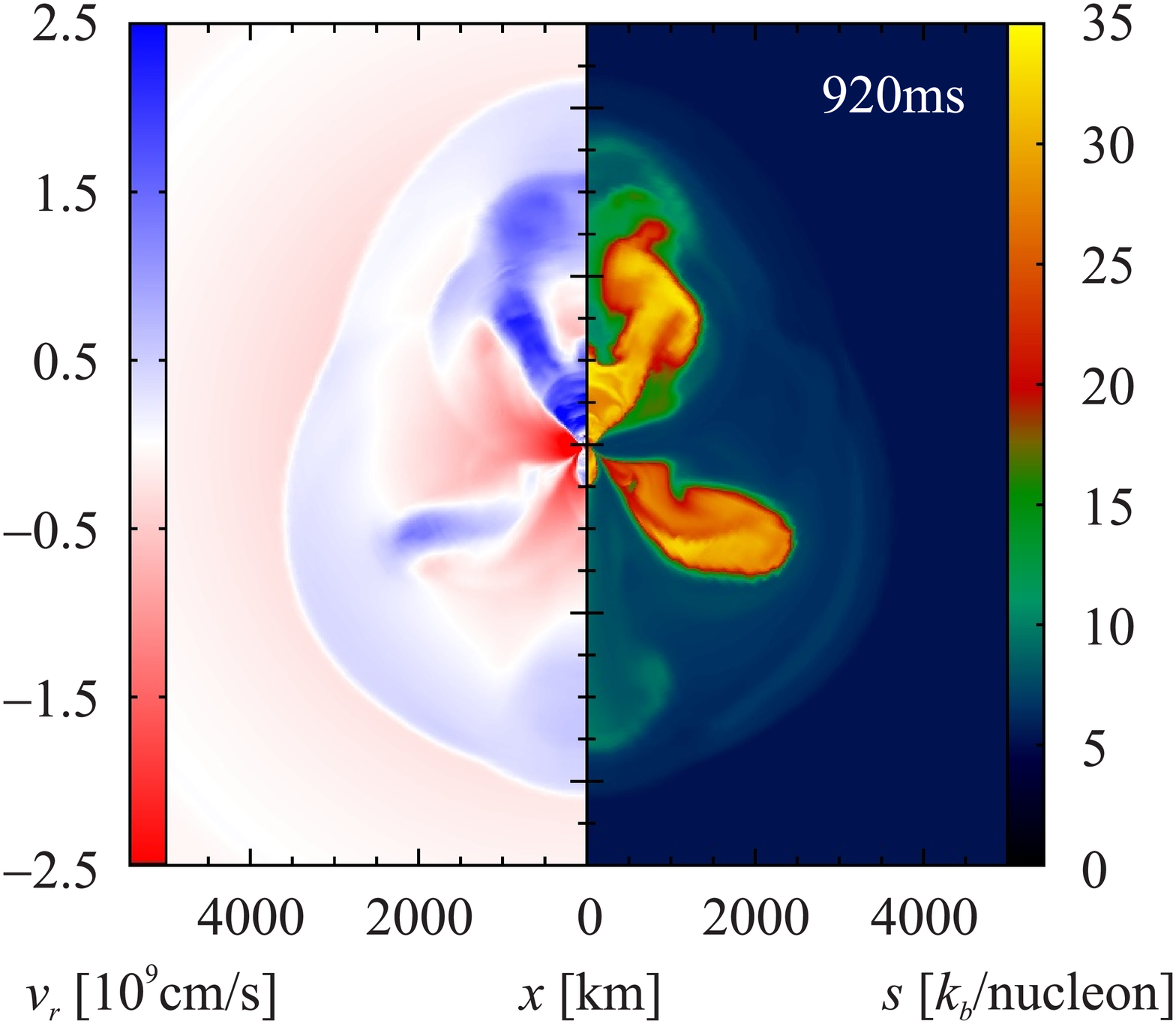}
\caption{Snapshots of the evolution of model G11, depicting the radial
  velocity $v_r$ (left half of panels) and the entropy per baryon $s$
  (right half of panels) $115 \ \mathrm{ms}$, $203 \ \mathrm{ms}$,
  $290 \ \mathrm{ms}$, $490 \ \mathrm{ms}$, $658 \ \mathrm{ms}$, and
  $920 \ \mathrm{ms}$ after bounce (from top left to bottom right).
\label{fig:s11_snapshots}
}
\end{figure*}

\begin{figure}
\plotone{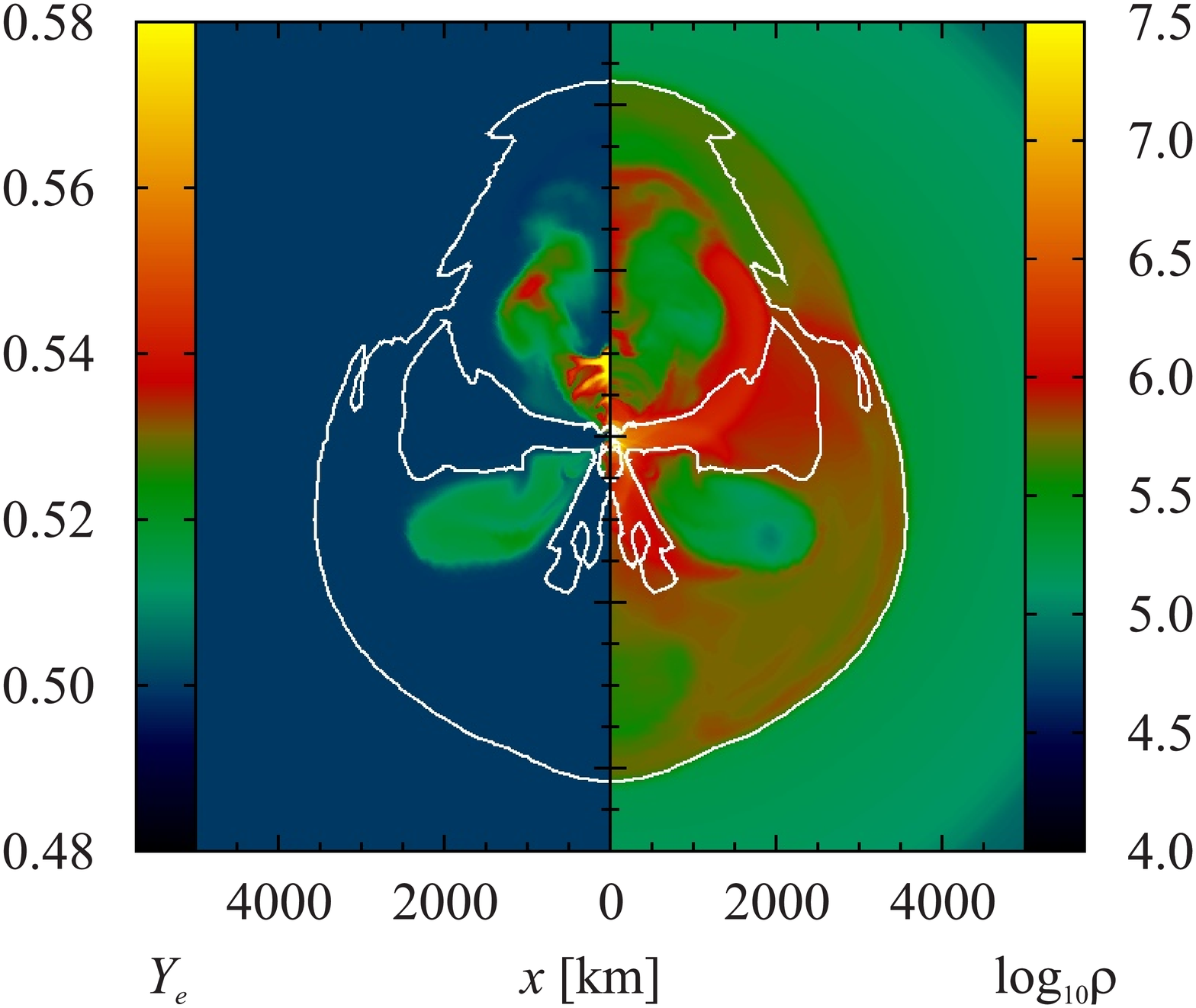}
\caption{Electron fraction (left half of panel) and density (right
  half of panel) at a time of $920 \ \mathrm{ms}$ after bounce in
  model G11. The white curves denote the boundary of the region where
  the local binding energy (Equation~\ref{eq:binding_energy}) is positive,
  i.e.\ they enclose the material that is preliminarily classified as
  ejecta and contains the high-$Y_e$ ($Y_e \approx 0.52$)
  bubbles of neutrino-heated matter.
\label{fig:ejecta_s11}
}
\end{figure}

Both the $11.2 M_\odot$ and the $15 M_\odot$ models show the
development of prompt post-shock convection a few milliseconds after
bounce, which then triggers early SASI activity at low amplitude for
$\approx 50 \ \mathrm{ms}$. This can be seen in
  Figure~\ref{fig:sasi}, where we show the normalized coefficients
  $a_\ell$ of the decomposition of the angle-dependent shock position
  $r_\mathrm{sh} (\theta)$ into Legendre polynomials up to
  $\l=3$. $a_\ell$ is computed as
\begin{equation}
a_\ell=
\frac{2 \ell + 1}{2}
\int\limits_{0}^\pi r_\mathrm{sh}(\theta) P_\ell \, \mathrm{d} \cos \theta.
\end{equation}

Hot-bubble convection then starts $\approx 70 \ldots 90 \ \mathrm{ms}$
after bounce and slowly pushes the shock further out than in 1D (see
Figure~\ref{fig:s15_shock_radius}). It again provides the seed for
fast growth of further SASI activity, but it is only at the time when
the Si/SiO interface reaches the shock that the SASI really starts to
become vigorous (Figure~\ref{fig:sasi}). Once these multi-dimensional
effects dominate the dynamics, the $11.2 M_\odot$ and the $15 M_\odot$
progenitor evolve rather differently.

\subsection{$11.2 M_\odot$ Model}
\subsubsection{Shock Propagation and Explosion Geometry}

In the case of the $11.2 M_\odot$ model, the Si/SiO-interface reaches
the shock $\approx 100 \ \mathrm{ms}$ after bounce, and the associated
drop in the mass accretion rate leads to strong shock expansion, thus
essentially enabling the approach to an the explosion as illustrated
by Figure~\ref{fig:entropy_poles}. During this phase, the post-shock
flow becomes dominated by the low-$\ell$-modes of the SASI
(Figure~\ref{fig:sasi}). As shown in Figure~\ref{fig:s11_snapshots}, the
buoyant convective plumes that contain neutrino-heated gas merge into
2--3 large bubbles. Around $280 \ \mathrm{ms}$ after bounce, the
tenuous polar plumes disappear almost completely for a short
while. This change of the flow geometry even results in a retraction
of the average shock radius for $70 \ \mathrm{ms}$ (left panel of
Figure~\ref{fig:shock_radius_and_explosion_energy}, red lines).  From
$350 \ \mathrm{ms}$ onward, however, the oscillations of the shock
become less violent as it is pushed steadily outward and
re-sphericized somewhat by sweeping up mass from the spherical
progenitor layers (left panels of Figure~\ref{fig:sasi} and
\ref{fig:shock_radius_and_explosion_energy}). This is also
  reflected by the normalized Legendre coefficients $a_\ell/a_0$ of
  the shock position, which decrease to a level of about $\sim 0.1$
  after $450 \ \mathrm{ms}$ post-bounce. From around $650
\ \mathrm{ms}$ onward, we see positive post-shock velocities along the
entire shock front.  Towards the end of the simulations, we observe
two high-entropy bubbles in the northern and southern hemisphere, a
rather broad downflow around the equatorial plane, and a further
downflow near the south polar axis. At this stage, the deformation of
the shock, expressed by the ratio of the maximum and minimum shock
radius is still appreciable:
$r_\mathrm{sh,max}/r_\mathrm{sh,min}\approx 4400 \ \mathrm{km}/ 3300
\ \mathrm{km} \approx 1.3$.  Even during the later stages of the
simulation, the geometry of the hot plumes does not freeze out, and
new downflows may still develop (cp.\ the snapshots at $490
\ \mathrm{ms}$ and $658 \ \mathrm{ms}$).  Accretion onto the
proto-neutron star therefore continues until late times, leading to a
growth of the baryonic mass of the proton-neutron star from $1.275
M_\odot$ to $1.36 M_\odot$ between $200 \ \mathrm{ms}$ and $920
\ \mathrm{ms}$.

Figure~\ref{fig:sasi} (left panel) shows that both the dipole
($\ell=1$) and quadrupole ($\ell=2$) mode are present in similar
strength in model G11, confirming that the deformation of the shock is
largest between $200 \ \mathrm{ms}$ and $450 \ \mathrm{ms}$ after
bounce and then decreases to a level of $\sim 0.1$ for the quadrupole
during the subsequent evolution.

\subsubsection{Explosion Energy}

We can compute a diagnostic ``explosion energy'' by integrating over
the material with positive binding energy $e_\mathrm{bind}$ at a
  certain time.  Since this energy does not account for subsequent
  nuclear recombination, burning, and the gravitational binding
  energy of the outer layers of the star, this quantity does not
  provide a direct measure for the final supernova explosion energy. 
In the GR case, we define $e_\mathrm{bind}$ in terms of the lapse
function $\alpha$, the rest-mass density $\rho$, the specific internal
energy $\epsilon$, the pressure $P$ and the Lorentz factor $W$ as
follows,
\begin{equation}
\label{eq:binding_energy}
e_\mathrm{bind}
=\alpha \left( \rho (c^2+\epsilon+P/\rho) W^2-P \right)-\rho W c^2.
\end{equation}
In order to maintain consistency with previous studies
\citep{buras_06_b,marek_09,bruenn_09}, we exclude rest-mass
contributions to the specific internal energy $\epsilon$.  It can
easily be verified that Equation~(\ref{eq:binding_energy}) correctly
reduces to
\begin{equation}
\label{eq:binding_energy_newton}
e_\mathrm{bind} \rightarrow \rho (\epsilon+\rho v^2/2+ \Phi)
\end{equation}
in the Newtonian limit (where $\Phi$ is the gravitational
potential)\footnote{Precisely speaking, we have $\alpha
  \rightarrow 1+\Phi/c^2$ and $W \rightarrow 1+v^2/2$ in the
  Newtonian limit, and obtain the Newtonian expression as an
  approximation to $\mathcal{O}(\epsilon/c^2,P/\rho c^2,v^2/c^2,\Phi/c^2)$.}
The diagnostic explosion energy is then computed by integrating over the
region where $e_\mathrm{bind}$ is positive,
\begin{equation}
E_\mathrm{expl}= \int\limits_{e_\mathrm{bind}>0} e_\mathrm{bind} \,\mathrm{d} \tilde{V}.
\end{equation}
Here, $\mathrm{d} \tilde{V}$ is the three-volume element for the curved
space-time metric (and not the flat-space volume element).

The time evolution of $E_\mathrm{expl}$ is plotted in the right panel
of Figure~\ref{fig:shock_radius_and_explosion_energy}, which shows that
material behind the shock first becomes nominally unbound $200
\ \mathrm{ms}$ after bounce for model G11. This corresponds to
  the time when the shock first expands beyond $\sim 400
  \ \mathrm{km}$, allowing the temperature to drop sufficiently for
  nucleon recombination to $\alpha$-particles to set in. The
diagnostic explosion energy slowly increases rather unsteadily at an average rate
of $6 \times 10^{49} \ \mathrm{erg} \ \mathrm{s}^{-1}$, and then seems
to level off around $3.5 \times 10^{49} \ \mathrm{erg}$ after $600 \ \mathrm{ms}$ post-bounce with some residual
fluctuations. By the end of simulation, the total mass of the
  material with positive binding energy amounts to $0.045 M_\odot$.

Despite this seeming ``saturation'' of $E_\mathrm{expl}$, no
definitive statement about the final explosion energy can be made as
yet at this stage, although we can tentatively estimate corrections
due to the energy input from nuclear burning in the shock,
$E_\mathrm{burn}$, the recombination of nucleons into
$\alpha$-particles (and possibly further into heavy nuclei), and the
deduction of the binding energy $E_\mathrm{preshock}$ of the unshocked
matter. All of these correction terms are of a similar magnitude as
the current diagnostic value $E_\mathrm{expl}$: 
The binding energy
$E_\mathrm{preshock}$ of \emph{all} the material ahead of the shock
is roughly $-7.5 \times 10^{49} \ \mathrm{erg}$
(i.e.\ exceeding $E_\mathrm{expl}$), which would have to be
included in the total explosion energy budget if these layers
were expelled completely. In reality, part of the pre-shock material
will not become unbound but accreted onto the proto-neutron star,
and the correction to the explosion energy will be smaller, but
only a (considerably) longer simulation could provide precise
values.

Recombination of nucleons and
$\alpha$-particles in the ejecta would provide an additional energy of
$E_\mathrm{rec}\approx 2 \times 10^{49} \ \mathrm{erg}$. Burning in
the shock will not yield any significant contribution with the current
shock velocities as the typical post-shock temperatures are already
too low ($<3 \times 10^9 \ \mathrm{K}$) to allow for explosive Si- and
O-burning.  The uncertainties in these numbers illustrate that in
order to obtain a reasonably accurate prediction for the explosion
energy of model G11, the simulation probably needs to be extended
until the shock reaches the C-O shell at $\gtrsim 2 \times 10^4
\ \mathrm{km}$ (at which point the binding energy of the remaining
pre-shock matter would be negligible). The final explosion energy
depends critically on the fraction of shocked material from this shell
that is accreted onto the proto-neutron star and thus need not become
unbound at all.

The slow growth and the stagnation of the diagnostic
explosion energy visible in
Figure~\ref{fig:shock_radius_and_explosion_energy} at late times is a
consequence of relatively inefficient neutrino heating in model G11
associated with an unfavorable 2D flow geometry in this case. In the
late phases, the rate of energy input by neutrinos in the gain region,
$\dot{Q}_\nu$, is only $\sim 5 \times 10^{50} \ \mathrm{erg}
\ \mathrm{s}^{-1}$. With a typical binding energy of a mass element at
the gain radius of $\sim 30 \ \mathrm{MeV} / \mathrm{baryon}$, this
implies that only $\sim 0.01 M_\odot \ \mathrm{s}^{-1}$ of additional
material can become unbound by neutrino heating. The actual mass flux
from the heating region into the ejecta is somewhat higher because
recombination of nucleons into $\alpha$-particles also contributes
part of the energy for unbinding the newly ejected material. Since the
ejecta from the gain region are channeled through relatively narrow
outflows into high-entropy bubbles
(cp.\ Figure~\ref{fig:s11_snapshots}) at high velocities, neutrino
heating is rather inefficient due to the short exposure time.  This
also implies that only a small excess energy -- i.e.\ much smaller
than the maximum energy available from recombination of $8.8
\ \mathrm{MeV} /\mathrm{baryon}$ -- remains for increasing the total
(internal+kinetic+gravitational) energy of the ejecta. Newly ejected
material thus adds only a few $10^{49} \ \mathrm{erg}
\ \mathrm{s}^{-1}$ to the explosion energy. However, the shock also
sweeps up material at a rate of about $0.05 M_\odot
\ \mathrm{s}^{-1}$, which implies a negative energy flux into the
``ejecta region'' of $\sim 5 \times 10^{49} \ \mathrm{erg}
\ \mathrm{s}^{-1}$. This may balance or even exceed the energy carried
by fresh ejecta from the gain region, thus accounting for the unsteady
evolution of $E_\mathrm{expl}$. With energy being fed into the ejecta
at such a low rate, a considerable fraction of the material swept up
by the shock will be channeled into the downflows and accreted onto
the proto-neutron star (Figure~\ref{fig:s11_snapshots}). We already
observe that the accretion rate onto the proto-neutron star starts to
increase again towards the end of the simulation, which leads to a
late-time rise of the electron neutrino and antineutrino luminosities
(which will be discussed in a subsequent paper on the neutrino
signal).  All these indications suggest that model G11 remains a
low-energy case and is likely to represent a fallback supernova,
i.e.\ the shock will propagate through the envelope and
  initially accelerate the swept-up material to positive velocities,
  but a large fraction of the shocked material will remain
  gravitationally bound and eventually fall back onto the neutron
  star.

\subsubsection{Ejecta Composition}
Although the final ejecta composition for our models can only be
determined by detailed nucleosynthesis calculations once the amount of
fallback is known, a few remarks can already be made about the
nucleosynthesis conditions on the basis of the entropy
(Figure~\ref{fig:s11_snapshots}) and the electron fraction
(Figure~\ref{fig:ejecta_s11}) of the material that is presumably
ejected.  Low-entropy matter ($s \lesssim 10 \ \mathrm{k}_b /
\mathrm{nucleon}$) with $Y_e \approx 0.5$ that has undergone little or
no neutrino heating, but has either been swept up directly by the
shock or has been pushed out by neutrino-heated bubbles before falling
inward to smaller radii, contributes most of the mass in the ``ejecta
region''. Depending on the maximum temperature reached before
expansion sets in, this material has partially been processed by
nuclear burning to various degrees: According to the simple
``flashing'' treatment in the \textsc{Vertex} code \citep{rampp_02},
intermediate-mass elements dominate the composition.

Hot, neutrino-processed material with entropies of up to $35
\ \mathrm{k}_b/\mathrm{nucleon}$ makes up only for a small fraction
 ($\sim$$0.005 M_\odot$ or $\sim$11\%) of the material
classified as ejecta by the end of the simulation. This part of the
ejecta is proton-rich with an electron fraction $Y_e$ ranging from
$\approx 0.51$ up to $\approx 0.58$. Different from the case of
electron-capture supernovae \citep{wanajo_11} and unlike
\citet{pruet_05}, who considered an artificially triggered 2D
explosion of the $15 M_\odot$ progenitor of \citet{woosley_95}, we do
not observe any slightly neutron-rich pockets in the ejecta, which is
a consequence of the different (slower) outflow dynamics (a detailed
analysis will be provided in a forthcoming paper). We therefore expect
the nucleosynthesis yields to conform with the established results for
proton-rich outflow conditions, i.e.\ the ejecta composition will be
dominated by ${}^{56} \mathrm{Ni}$ and helium with an admixture of a
few rare isotopes (${}^{45} \mathrm{Sc}$, ${}^{49} \mathrm{Ti}$, and
${}^{64} \mathrm{Zn}$) with large production factors
\citep{pruet_05,froehlich_06a}.  Depending on the neutrino
luminosities there may also be potential for $\nu p $-process
nucleosynthesis \citep{froehlich_06b,pruet_06}.

\subsection{$15 M_\odot$ Model}

\begin{figure*}
\plottwo{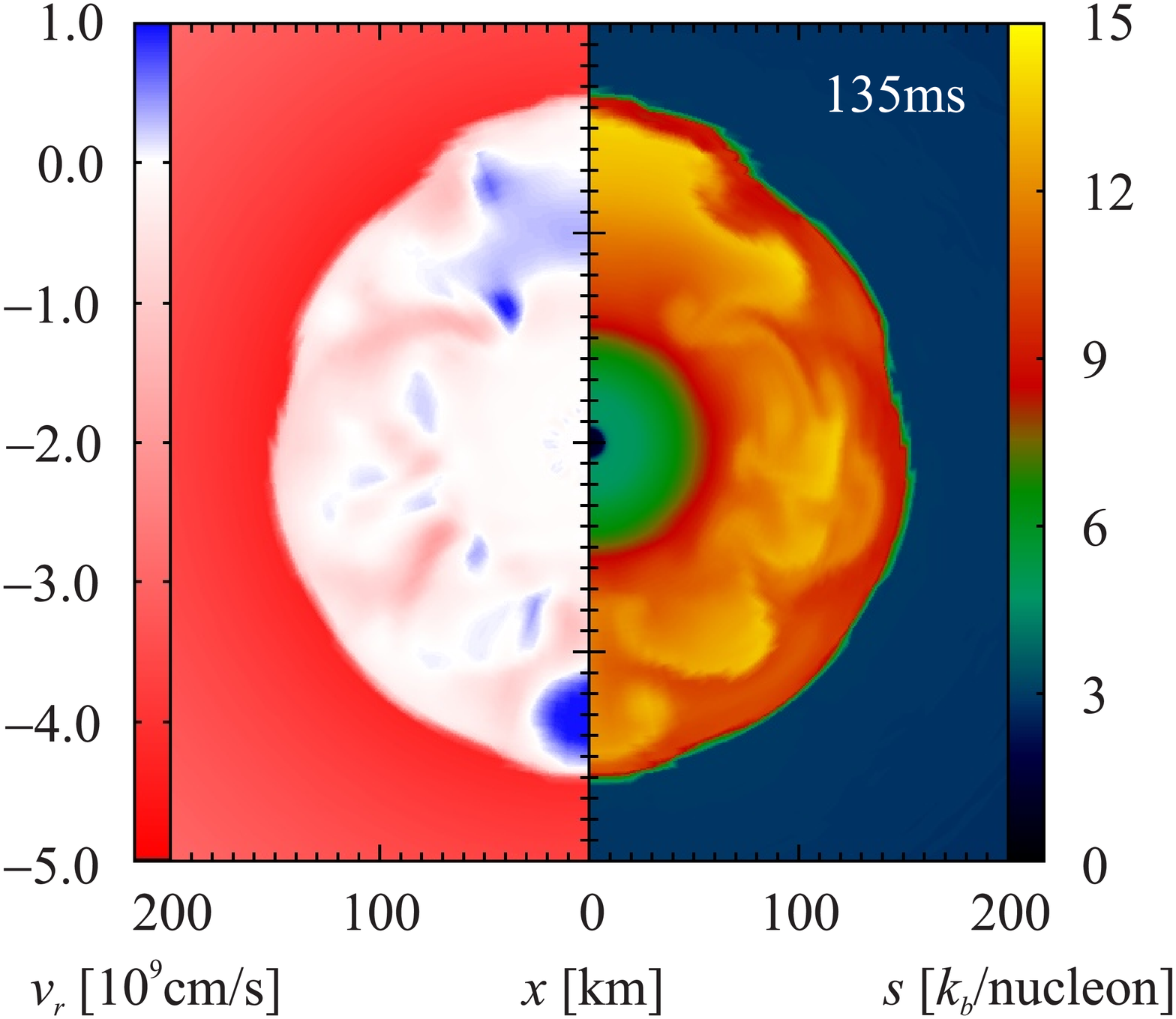}{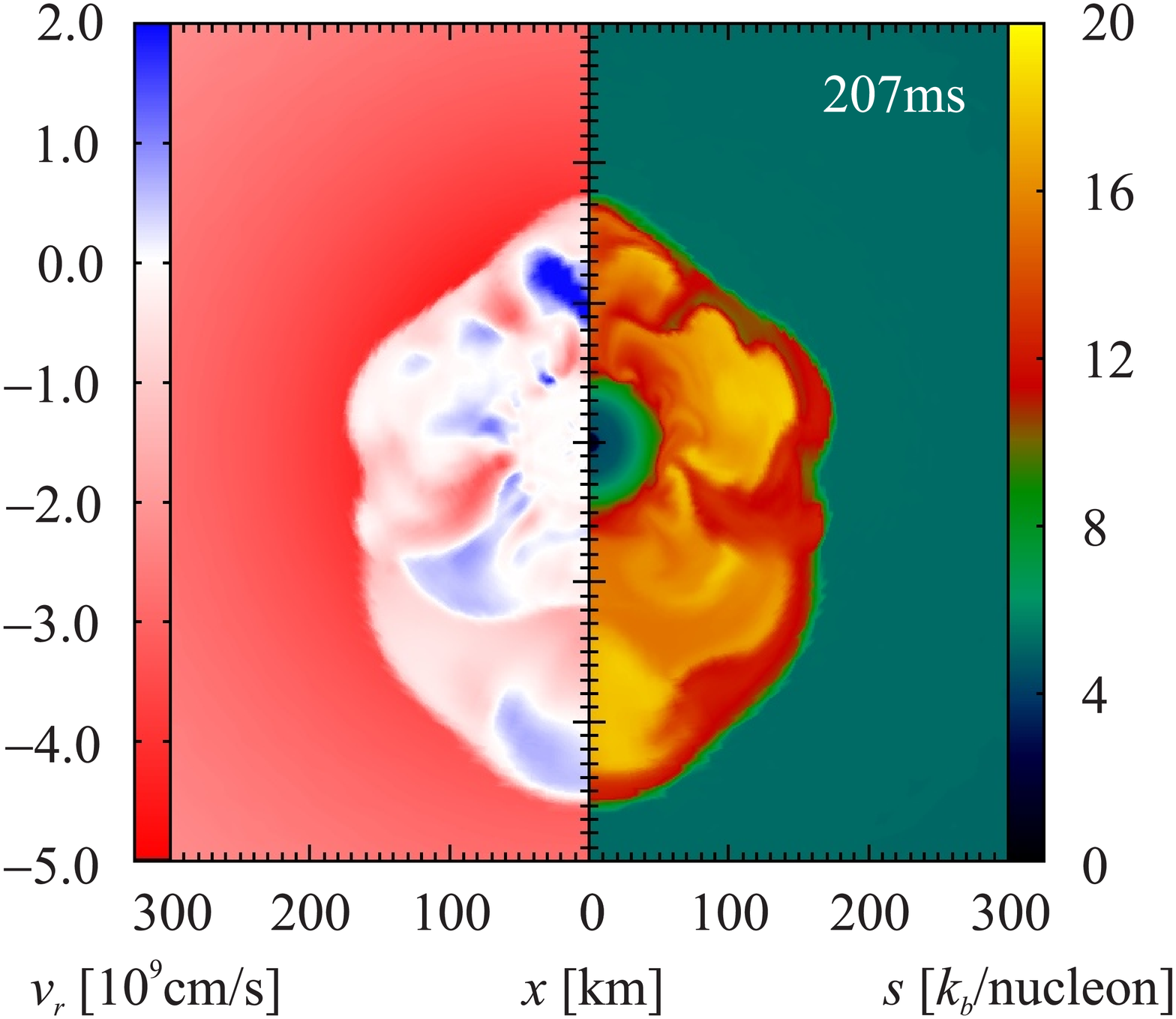}
\vspace{0.2in}
\plottwo{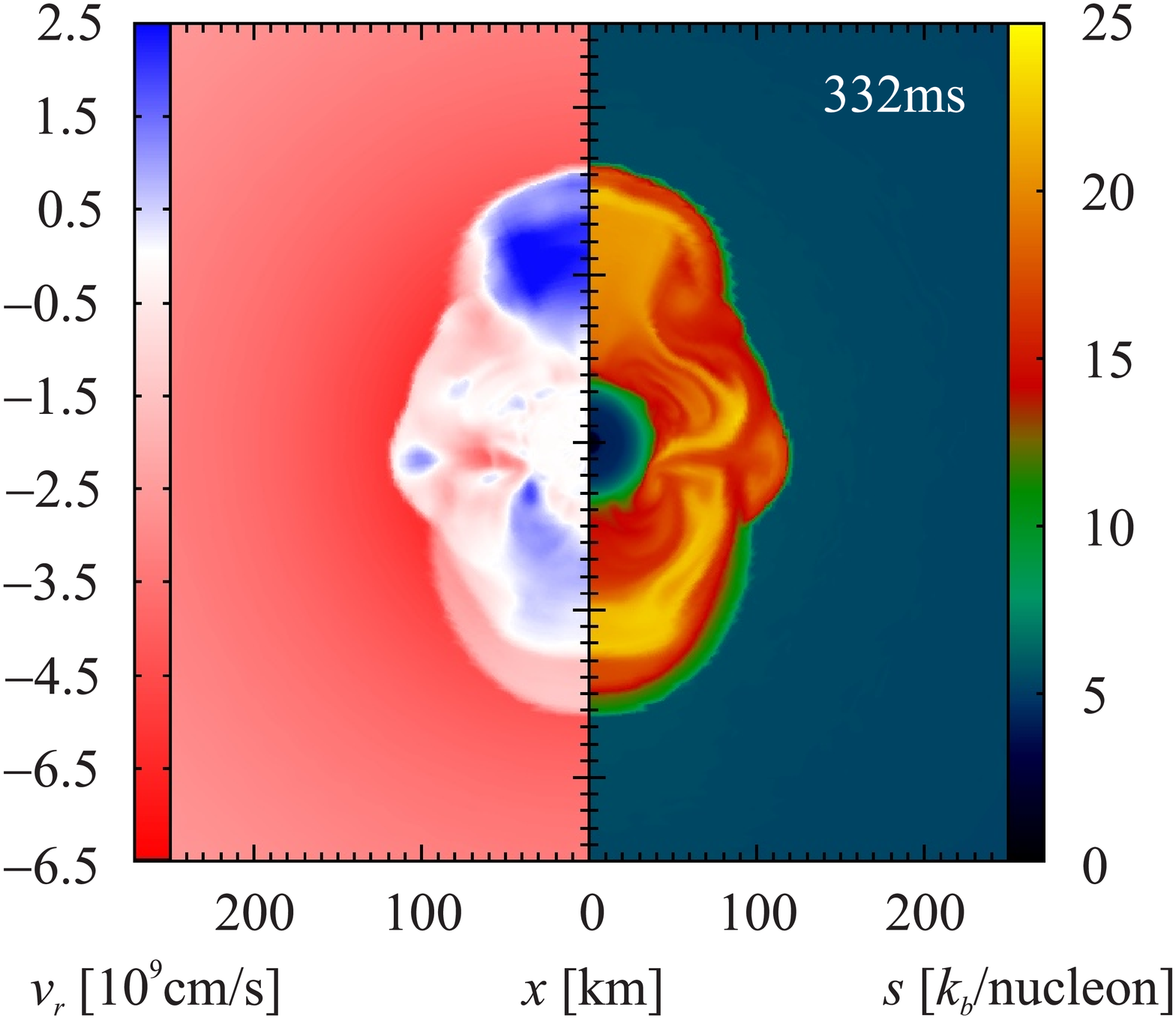}{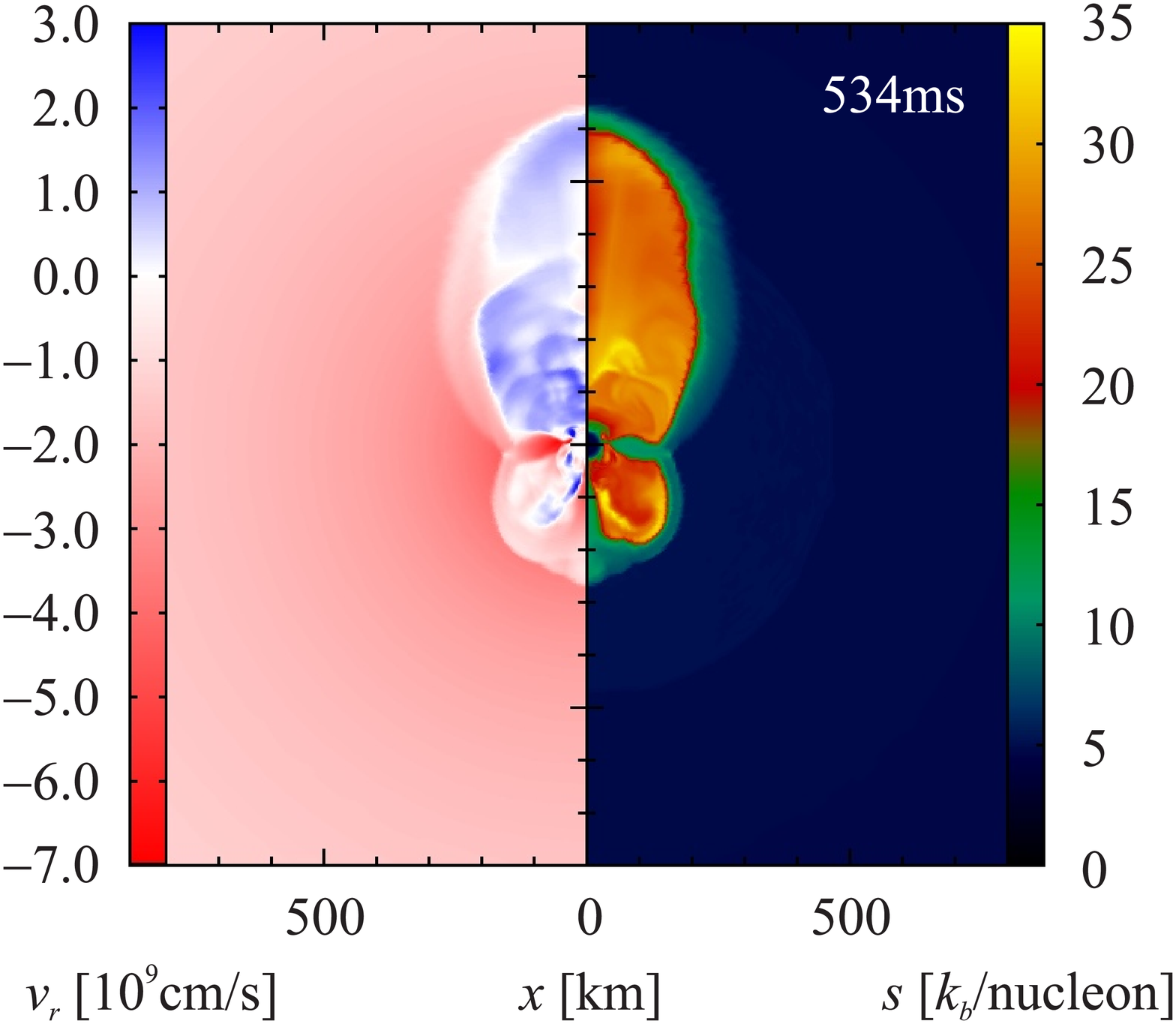}
\vspace{0.2in}
\plottwo{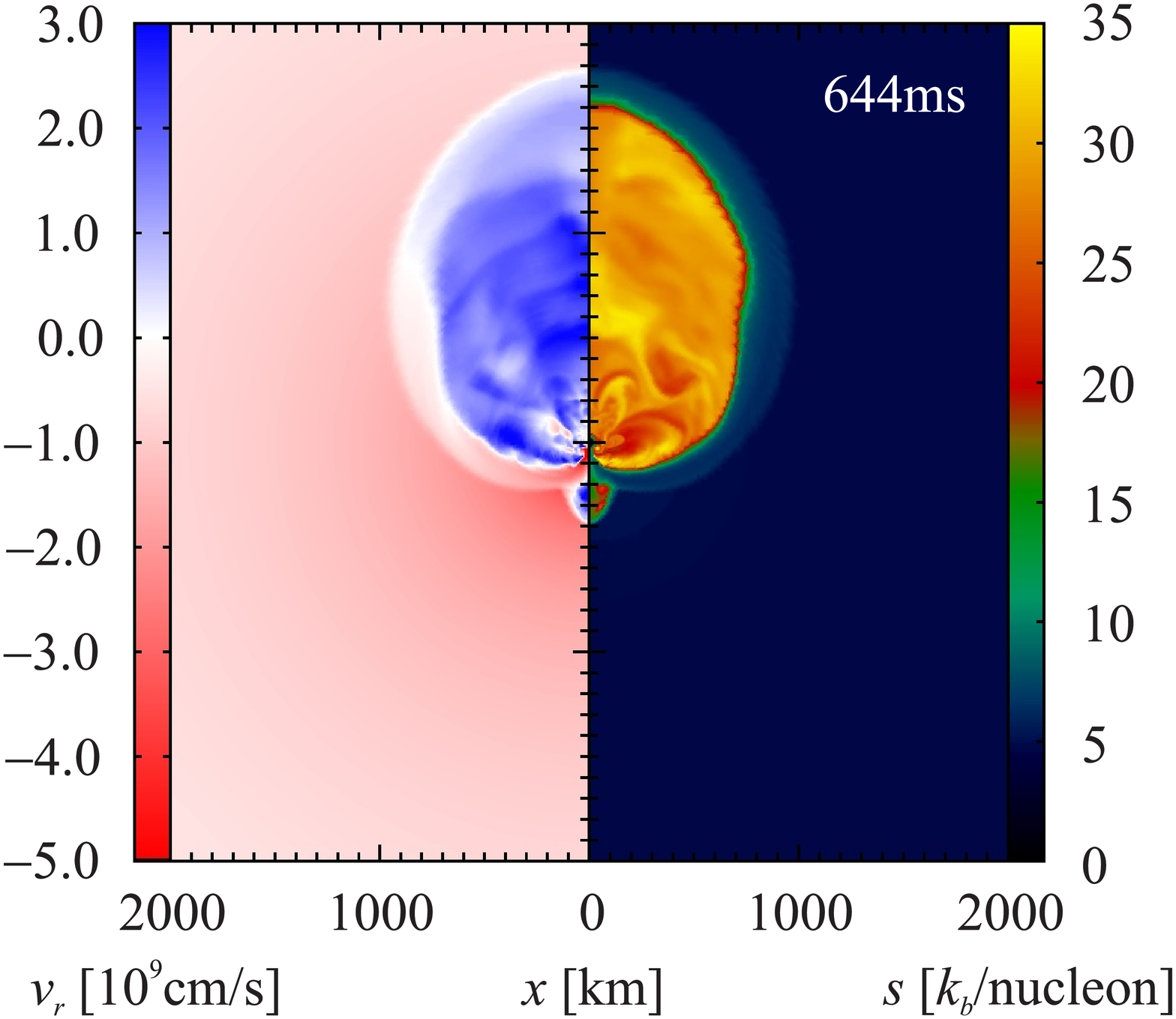}{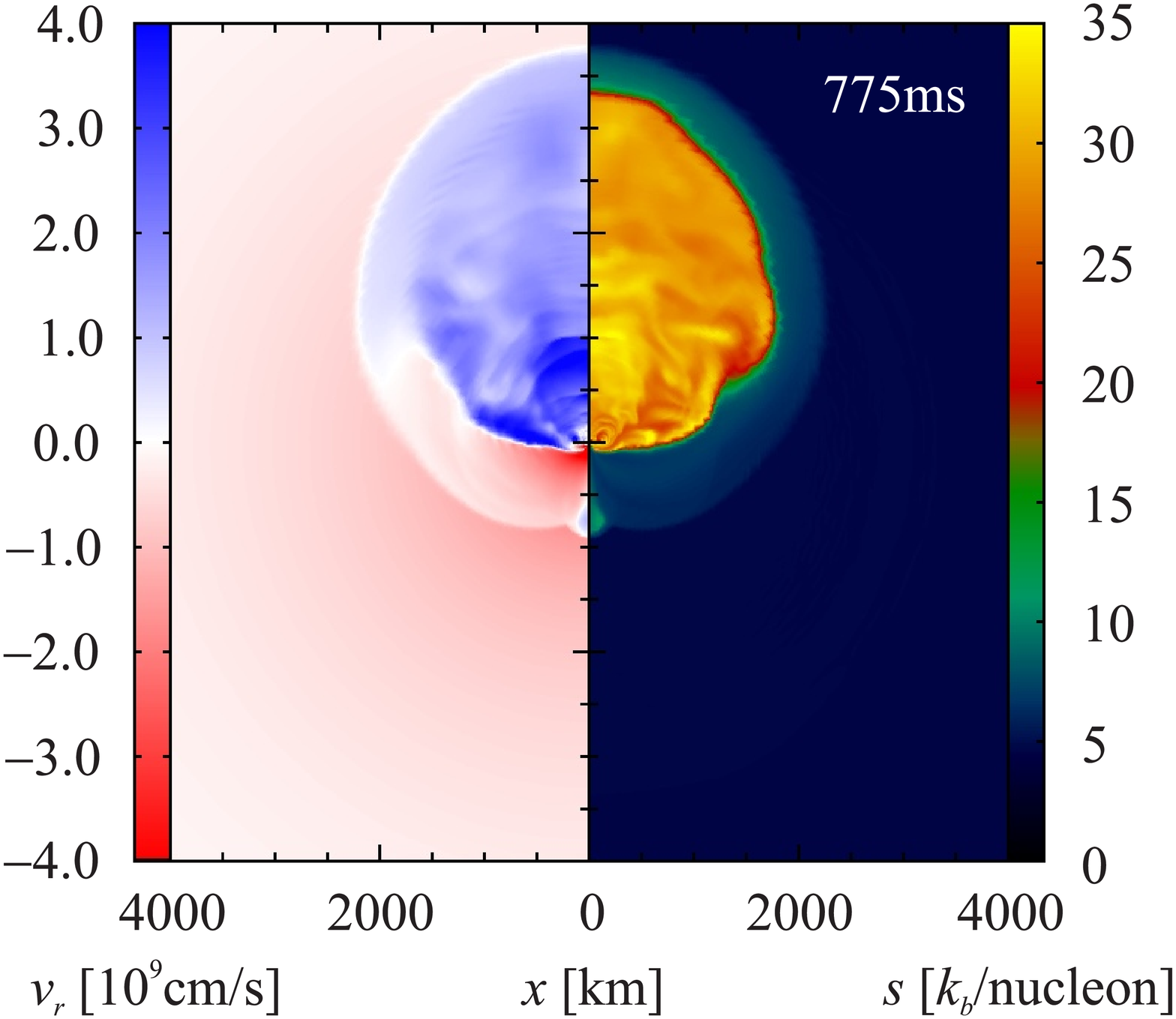}
\caption{Snapshots of the evolution of model G15, depicting the radial
  velocity $v_r$ (left half of panels) and the entropy per baryon $s$
  (right half of panels) $135 \ \mathrm{ms}$, $207 \ \mathrm{ms}$,
  $332 \ \mathrm{ms}$, $534 \ \mathrm{ms}$, $644 \ \mathrm{ms}$, and
  $775 \ \mathrm{ms}$ after bounce (from top left to bottom right).
\label{fig:s15_snapshots}
}
\end{figure*}

\begin{figure}
\plotone{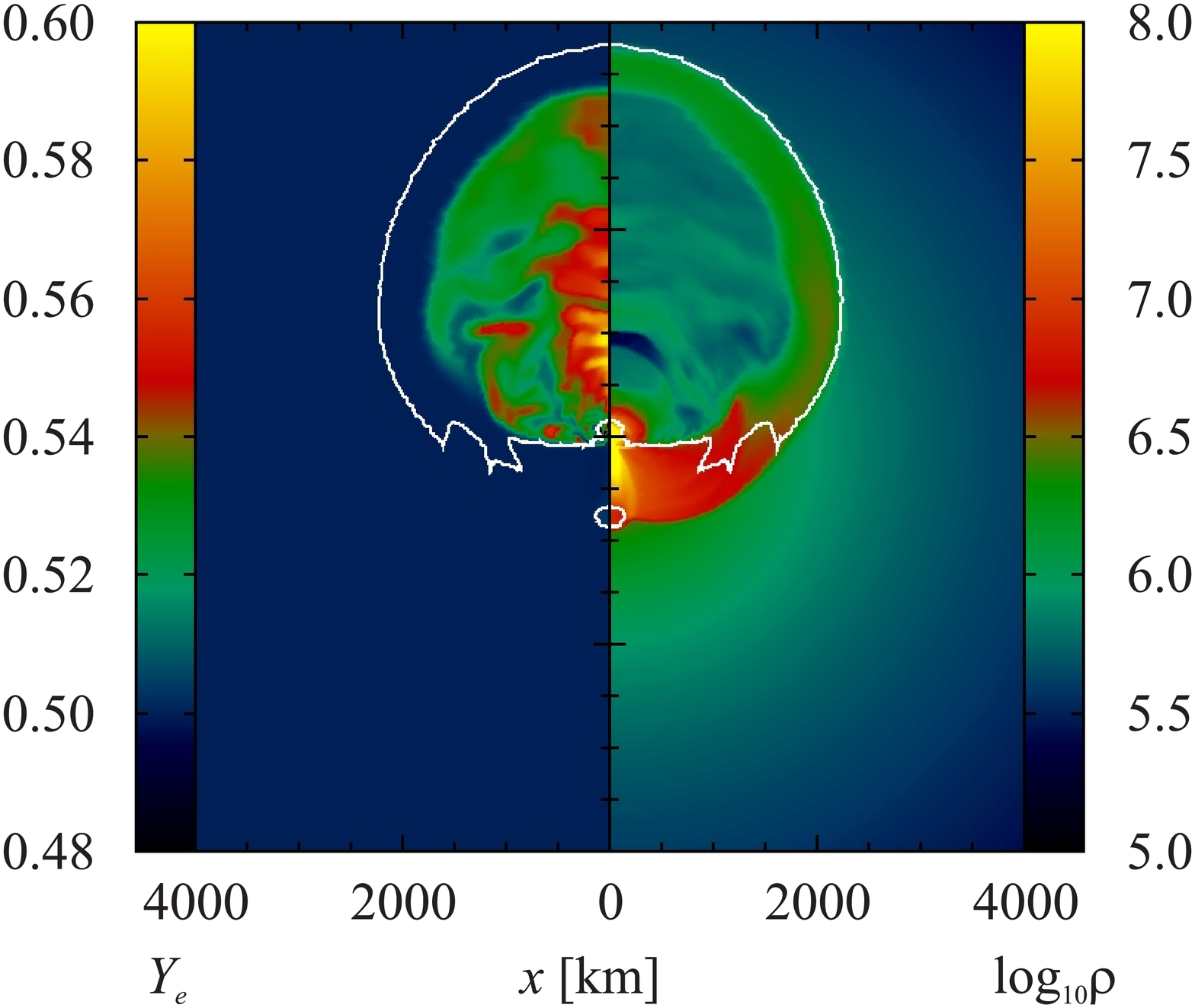}
\caption{Electron fraction (left half of panel) and density (right
  half of panel) at a time of $775 \ \mathrm{ms}$ after bounce for
  model G15. The white curves denote the boundary of the region where
  the local binding energy (Equation~\ref{eq:binding_energy}) is positive,
  i.e.\ they enclose the material that is preliminarily unbound and classified as
  ejecta (mostly located in the northern hemisphere) and contains the
  high-$Y_e$ ($Y_e \approx 0.52$) bubbles of neutrino-heated matter.
\label{fig:ejecta_s15}
}
\end{figure}

\subsubsection{Shock Propagation and Explosion Morphology}

Different from the $11.2 M_\odot$ case, the development of the
explosion in model G15 is not immediately connected to the Si/SiO
interface reaching the shock although the decrease in the accretion
rate results in a transient increase of the average shock to $220
\ \mathrm{km}$ at $200 \ \mathrm{ms}$ after bounce (see
Figure~\ref{fig:s15_shock_radius} and Figure~\ref{fig:entropy_poles})
and also in increased SASI activity (Figure~\ref{fig:sasi}, right
panel). However, $230 \ \mathrm{ms}$ after bounce the shock again
starts to recede slowly with the average shock radius reaching a
minimum value of about $100 \ \mathrm{km}$ at $380
\ \mathrm{ms}$. During this period, the SASI remains active with
strong dipole and quadrupole components (the maximum amplitudes being
$a_1/a_0 \approx a_2/a_0 \approx 0.3$, Figure~\ref{fig:sasi}, right
panel). Around $400 \ \mathrm{ms}$, the average shock radius begins to
move outward rather steadily (Figure~\ref{fig:s15_shock_radius}), and
at about $430 \ \mathrm{ms}$, some material becomes nominally unbound
(Figure~\ref{fig:shock_radius_and_explosion_energy}). Model G15
develops a strongly asymmetric explosion
(Figures~\ref{fig:entropy_poles},\ref{fig:shock_radius_and_explosion_energy},\ref{fig:s15_snapshots}):
By the end of the simulation, the shock has reached $3800
\ \mathrm{km}$ in the northern hemisphere, while the minimum shock
radius over the only remaining strong downflow in the southern
hemisphere is only $850 \ \mathrm{km}$
(Figure~\ref{fig:shock_radius_and_explosion_energy}); i.e.\ the ratio
$r_\mathrm{max}/r_\mathrm{min}$ of the maximum and minimum shock
radius is as large as 4.5:1. Snapshots of the developing asymmetric
explosion with even more extreme shock deformation during earlier
phases of the explosion are shown in Figure~\ref{fig:s15_snapshots}.

\subsubsection{Explosion Energy}

From $540 \ \mathrm{ms}$ onward, we observe a much more rapid growth
of the diagnostic explosion energy than for model G11. By $770 \ \mathrm{ms}$, it
has reached a value of $E_\mathrm{expl} = 1.3 \times 10^{50}$ and
still continues to increase at a rate of $\approx 7\times 10^{50}
\ \mathrm{erg} \ \mathrm{s}^{-1}$. This rate of increase is roughly
consistent with the assumption that neutrino heating at a rate of $(4
\ldots 5) \times 10^{51} \ \mathrm{erg} \ \mathrm{s}^{-1}$
(Figure~\ref{fig:gain_region_comparison}) allows an outflow rate from
the gain region of $\sim 0.06 M_\odot \ \mathrm{s}^{-1}$ (assuming a
binding energy of $\sim 40 \ \mathrm{MeV} / \mathrm{baryon}$ at the
gain radius), and that recombination of nucleons into
$\alpha$-particles leads to an excess energy of $\sim 8 \ \mathrm{MeV}
/ \mathrm{baryon}$ that actually contributes to the explosion energy.
As for model G11, more material is actually swept up by the shock
($\sim 0.15 M_\odot \ \mathrm{s}^{-1}$) than is supplied from the
heating region, but in contrast to model G11, the corresponding
negative total energy flux into the ejecta region is much smaller
($\sim 10^{50} \ \mathrm{erg} \ \mathrm{s}^{-1}$) than the positive
contribution from hot, neutrino-heated ejecta ($\sim 10^{51}
\ \mathrm{erg} \ \mathrm{s}^{-1}$).  $E_\mathrm{expl}$ therefore
stably increases for model G15.  The presence of a strong and stable
downflow in the southern hemisphere is probably helpful for this
behavior: Compared to model G11, the accretion rate onto the
proto-neutron star is considerably higher ($\dot{M} \sim 0.1 M_\odot
\ \mathrm{s}^{-1}$) during the late phases, which results in a sizable
accretion contribution to the neutrino luminosity and therefore allows
for persistently strong neutrino heating in the gain region. Due
to continuous accretion, the baryonic mass of the PNS has grown
to $1.58 M_\odot$ by the end of the simulation.

With respect to the final explosion energy, the same reservations
apply as for model G11, i.e.\ the binding energy of the pre-shock
material (roughly $2.6 \times 10^{50} \ \mathrm{erg}$) is still larger
than the current value of $E_\mathrm{expl}$. However, different from
model G11 we expect a significant energy input from burning in the
shock on the order of $(1 \ldots 2) \times 10^{50} \ \mathrm{erg}$ or
more, i.e.\ model G15 has almost reached the stage where the envelope
could become unbound completely even without further neutrino heating.

\subsubsection{Ejecta Composition}
The composition of the ejecta in model G15 exhibits some marked
differences to model G11, although we can identify the same two
components: low-entropy matter with $Y_e=0.5$ swept up by the shock, and
proton-rich, neutrino-heated matter
(Figures~\ref{fig:s15_snapshots},\ref{fig:ejecta_s15}). Different from
model G11, much of the shocked material accumulated by the large
expanding bubble in the northern hemisphere has been burnt to $^{56}
\mathrm{Ni}$, and even at $775 \ \mathrm{ms}$ post-bounce, the
post-shock temperature is still high enough to allow at least for explosive
O-burning.  Moreover, hot, neutrino-heated material accounts for a
larger fraction of the ejecta ($0.01 M_\odot$, or roughly one third of
the matter with positive total energy). We again find the
neutrino-heated ejecta to be exclusively proton-rich, but both the
typical and the maximum electron fraction in the neutrino-heated
ejecta are even higher than for model G11 with $Y_e$ ranging up to
$0.6$.  

\section{Analysis of Heating Conditions and Comparison to
Non-Exploding Models}
\label{sec:heating}
It is interesting to compare the evolution of model G15 to the
pseudo-Newtonian run M15, the purely Newtonian model N15, and to
simulation S15 with simplified neutrino reaction rates.  The average
shock radius is relatively similar in all cases until $\approx 400
\ \mathrm{ms}$ after bounce (Figure~\ref{fig:s15_shock_radius}),
except for model N15, which maintains a larger average shock radius
(by $30 \ldots 50 \ \mathrm{km}$). Among G15, M15, and S15, the shock
in model G15 reacts most strongly to the drop in $\dot{M}$ associated
with the Si/SiO interface and stays a little further out until $380
\ \mathrm{ms}$, but the differences between the three models remain
rather modest; right before the onset of the explosion in model G15,
the shock position is virtually identical. Yet, only a few tens of
milliseconds later, an explosion develops in model G15, whereas S15,
M15 and even N15 with its relatively large shock radius ($\approx 140
\ \mathrm{km}$) show no sign of shock revival. In this section, we
discuss the reason for this (unexpectedly) different behavior of the
models lacking either the GR hydrodynamics treatment or the full set
of neutrino opacities.

\subsection{Analysis Framework}
Shock revival can be understood in terms of a competition of neutrino
heating and the downward advection of the gas in the gain layer
(e.g.\ \citealp{burrows_93,janka_01}). A detailed analysis shows that
there is a critical neutrino luminosity for each value of the mass
accretion rate $\dot{M}$ of the shock, above which no steady-state
accretion solution can exist
\citep{burrows_93,janka_01,murphy_08b,pejcha_11,fernandez_12}.
  Roughly speaking, conditions become favorable for an explosion if
  the time the accreted material spends in the gain region (measured
  by the ``advection'', ``residency'', or ``dwelling''
  time-scale $\tau_\mathrm{adv}$) becomes longer than the time
  required to lift its binding energy to positive values by neutrino
  heating (the heating time-scale $\tau_\mathrm{heat}$). If a ratio
  $\tau_\mathrm{adv} / \tau_\mathrm{heat} \gtrsim 1$ is maintained for
  a sufficiently long time (typically a few tens of milliseconds), the
  shock can, in all probability, expand sufficiently to create a
  positive feedback loop by further increasing $\tau_\mathrm{adv}$,
  thus establishing a runaway situation with sustained shock expansion
  that eventually leads to an explosion. This concept provides an
  adequate basis for understanding the different outcome of the
  exploding models G11 and G15 in contrast to the other $15 M_\odot$
  runs (cp.\ \citealp{marek_09}).

We compute $\tau_\mathrm{adv}$ as the ratio of the binding
energy $|E_\mathrm{gain}|$ of the material in the gain region
and the volume-integrated energy deposition rate in that region
(in agreement with \citealt{marek_09} and with the best 
definition identified by \citealt{fernandez_12}),
\begin{equation}
\tau_\mathrm{heat}=\frac{|E_\mathrm{gain}|}{\dot{Q}_\mathrm{heat}}.
\end{equation}
Here, $E_\mathrm{gain}$ and $\dot{Q}_\mathrm{heat}$ are
volume-integrals over the binding energy density $e_\mathrm{bind}$
as given by Equation~(\ref{eq:binding_energy})
and the local neutrino heating rate per unit volume $\dot{q}_e$ between
the gain radius $r_\mathrm{gain}$ (computed from
the angle-averaged neutrino heating profile) and the (average) shock radius
$r_\mathrm{shock}$,
\begin{equation}
E_\mathrm{gain}=
\int\limits_{r_\mathrm{gain} < r < r_\mathrm{shock}}
e_\mathrm{bind} \mathrm{d} \tilde{V},
\end{equation}
\begin{equation}
\dot{Q}_\mathrm{heat}=
\int\limits_{r_\mathrm{gain} < r < r_\mathrm{shock}}
\dot{q}_e \mathrm{d} \tilde{V}.
\end{equation}
There are alternative definitions for the advection time-scale
$\tau_\mathrm{adv}$ \citep{buras_06_b,marek_09,murphy_08b,pejcha_11},
each of which can be supported by very plausible arguments.  For the sake of
convenience, we choose to express it in terms of two readily available
quantities, the accretion rate $\dot{M}$ of gas through the shock and
the baryonic rest mass contained in the gain layer $M_\mathrm{gain}$,
\begin{equation}
\tau_\mathrm{adv}=M_\mathrm{gain}/\dot{M},
\end{equation}
where $M_\mathrm{gain}$ is given by
\begin{equation}
M_\mathrm{gain}=
\int\limits_{r_\mathrm{gain} < r < r_\mathrm{shock}}
\rho W \mathrm{d} \tilde V.
\end{equation}
This definition measures the time matter needs to flow through the
gain region if steady-state conditions hold, and is used here because
it allows for a very straightforward evaluation of
$\tau_\mathrm{adv}/\tau_\mathrm{heat}$ not only in the 1D case,
  but also in the multi-dimensional case.  The time-scale ratio
  $\tau_\mathrm{adv}/\tau_\mathrm{heat}$ was recently also shown by
  \citet{fernandez_12} to provide a useful instrument for
distinguishing models that are going to explode from ``pessimistic'' ones.

\subsection{$11.2 M_\odot$ Progenitor}

\begin{figure}
\plotone{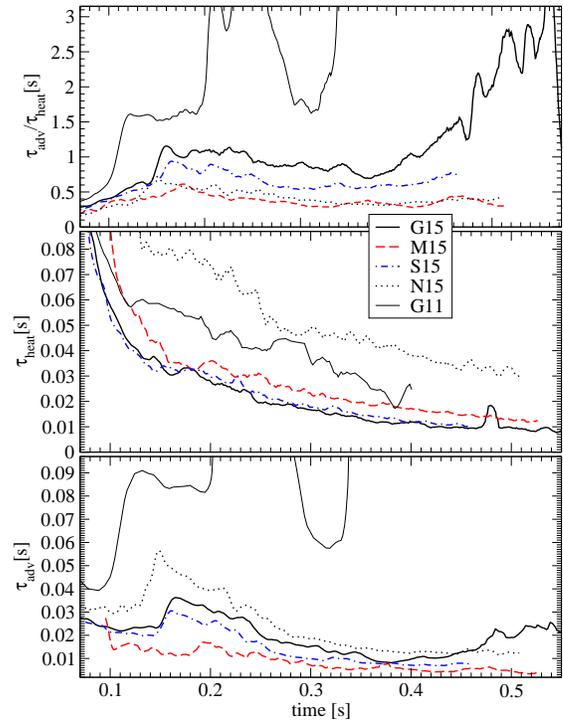}
\caption{Runaway criterion $\tau_\mathrm{adv} / \tau_\mathrm{heat}$,
  heating time-scale $\tau_\mathrm{heat}$, and advection time-scale
  $\tau_\mathrm{adv}$ for models G15 (thick black solid line), M15
  (thick, red, dashed), S15 (thick, blue, dash-dotted), N15 (thick,
  black, dotted), and G11 (thin solid line, black) smoothed over $10
  \ \mathrm{ms}$. Note that we do not evaluate the time-scales during
  the later phase of the explosion when the their definitions are no
  longer meaningful. Naturally, the time-scales cannot be computed
  before the formation of the gain layer several tens of milliseconds
  after bounce either.
\label{fig:timescale_comparison}
}
\end{figure}

The evaluation of the time-scale ratio $\tau_\mathrm{adv} /
\tau_\mathrm{heat}$ confirms that the transition to shock expansion of
model G11 is a relatively clear-cut case, and is indeed triggered by
the Si/SiO-interface. As shown by Figure~\ref{fig:timescale_comparison},
the expansion of the shock due to the drop in the mass accretion rate
results in a considerable increase of the advection time-scale
$\tau_\mathrm{adv}$ (by a factor of $2 \ldots 3$) within a few tens of
milliseconds, which is sufficient to bring the critical time-scale
ratio $\tau_\mathrm{adv} / \tau_\mathrm{heat}$ above unity in
conjunction with the support from hot-bubble convection. At later
times, $\tau_\mathrm{adv} / \tau_\mathrm{heat}$ never drops below
unity, not even during the transient phase of shock retraction around
$300 \ \mathrm{ms}$, indicating heating conditions that favor robust
shock expansion. These findings are in very good agreement with the
pseudo-Newtonian model of \citet{buras_06_b} and \citet{marek_09},
even though the explosion morphology is different (spherical instead
of dipolar), which is probably the result of stochastic variations of
the SASI activity.

\begin{figure}
\plotone{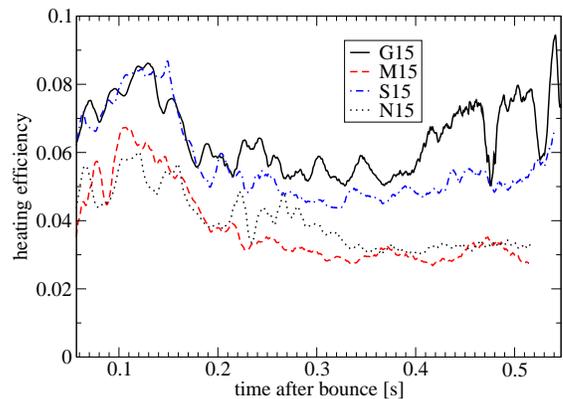}
\caption{Heating efficiency for models G15 (black solid line), M15
  (red, dashed), S15 (blue, dash-dotted), and N15 (black, dotted). The
  heating efficiency is computed as the ratio
  $\dot{Q}_\mathrm{heat}/(L_{\nu_e}+L_{\bar{\nu}_e})$ of the
  volume-integrated neutrino heating rate $\dot{Q}_\mathrm{heat}$ and
  the sum of the $\nu_e$ and $\bar{\nu}_e$ luminosities.
\label{fig:heating_efficiency}}
\end{figure}

\begin{figure*}
\plotone{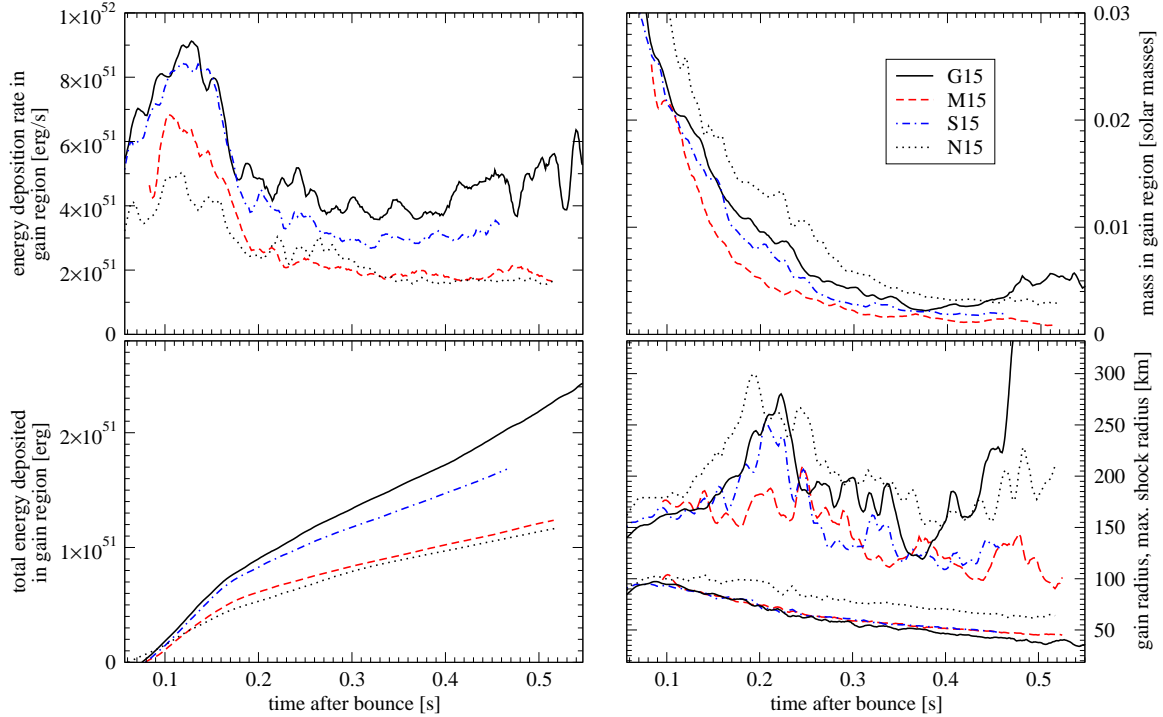}
\caption{
\label{fig:gain_region_comparison}
The energy deposition rate in the gain region (top left panel,
smoothed over $10 \ \mathrm{ms}$), the total (time-integrated) energy
deposited in the gain region (bottom left), the total mass in the gain
region (top right, smoothed over $10 \ \mathrm{ms}$), the gain radius
(bottom right, upper set of lines) and the maximum shock radius
(bottom right, lower set of lines)} for models G15 (black solid
line), M15 (red, dashed), S15 (blue, dash-dotted), and N15 (black,
dotted). 
\end{figure*}

\begin{figure}
\plotone{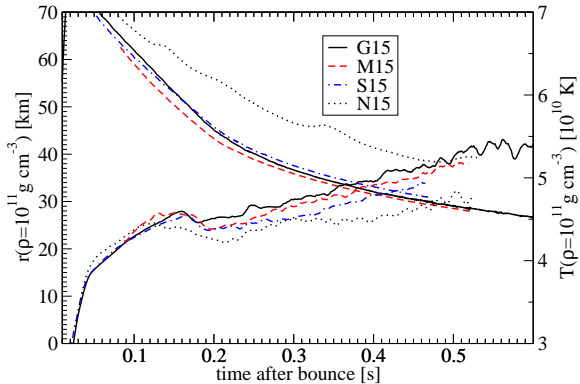}
\caption{Circumferential radius (decreasing with time) and surface
  temperature (increasing) of the proto-neutron star for models G15,
  M15 S15, and N15. The surface is defined by a fiducial density value
  of $\rho=10^{11} \ \mathrm{g} \ \mathrm{cm}^{-3}$.
\label{fig:pns_radius_and_temperature}
}
\end{figure}

\begin{figure*}
\plotone{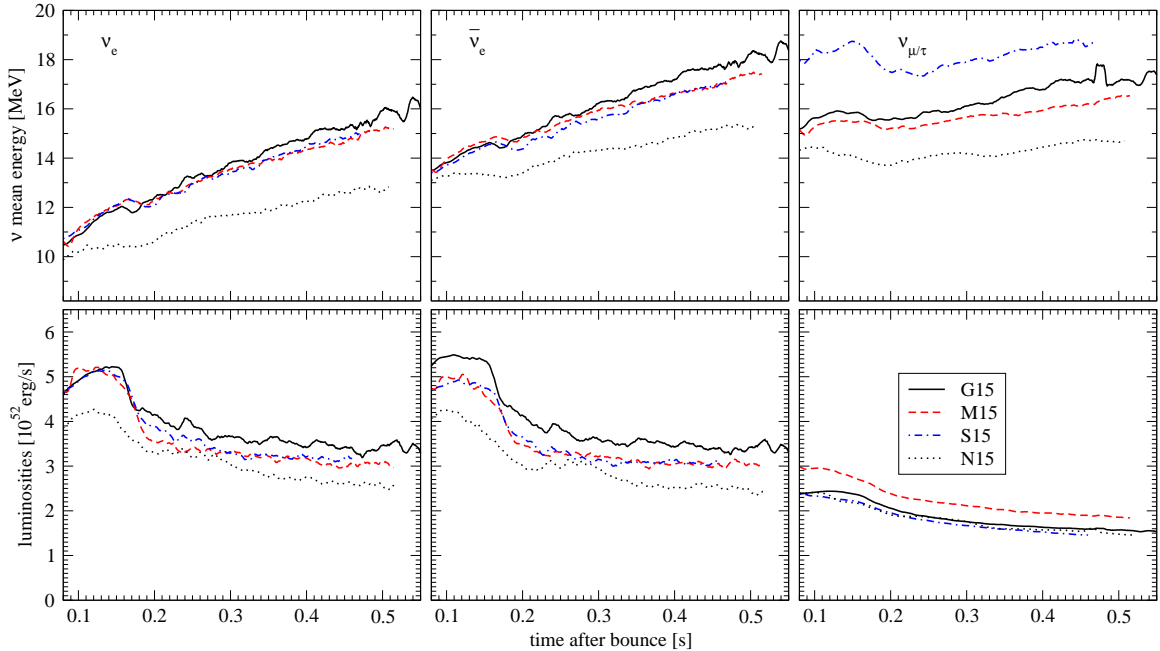}
\caption{
\label{fig:neutrino_gain_comparison}
Neutrino mean energies (upper panels) and luminosities (lower panels)
at the gain radius for the relativistic model G15 (black solid lines),
the pseudo-Newtonian model M15 (red dashed lines), the relativistic
model S15 with simplified neutrino rates (blue dash-dotted lines), and
the purely Newtonian run N15 (black, dotted). Electron neutrinos,
electron antineutrinos, and $\mu/\tau$ neutrinos are shown in the
left, middle, and right panel of each row, respectively. The mean
energy is defined as the ratio of the angle-averaged neutrino energy
and number density, and the luminosity is computed as the integral of
the flux over the sphere corresponding to the gain radius. The
luminosity given here is for a \emph{single} species of
$\nu_{\mu/\tau}$, not for all four of them.}
\end{figure*}

\begin{figure*}
\plottwo{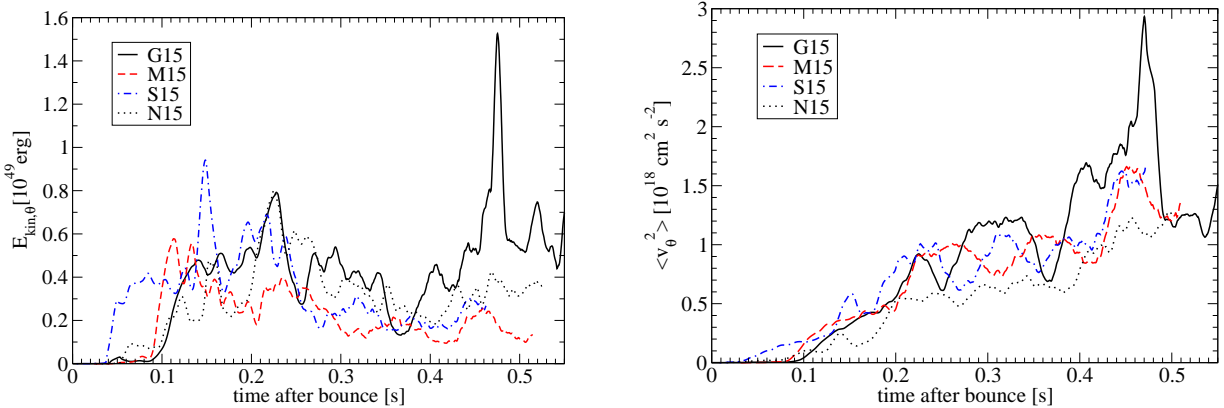}{f15b.eps}
\caption{Left: The kinetic energy $E_{\mathrm{kin},\theta}$ associated
  with lateral motions in the gain region for models G15 (black solid
  line), M15 (red, dashed), S15 (blue, dash-dotted), and N15 (black,
  dotted); the curves are smoothed over $10
  \ \mathrm{ms}$. $E_{\mathrm{kin},\theta}$ can be taken as a possible
  measure for the combined activity of convection and the SASI, which
  both contribute to this quantity. Note that a comparison of
  different models is difficult due to the fact that
  $E_{\mathrm{kin},\theta}$ does not only depend on the typical velocities
  but also on the mass initially available in the gain region.  Right:
  The velocity dispersion $\left \langle v_\theta^2 \right \rangle$ in
  the gain region, which is related to $E_{\mathrm{kin},\theta}$ by
  the relation $\left \langle v_\theta^2 \right \rangle = 2
  E_{\mathrm{kin},\theta}/M_\mathrm{gain}$.  $\left \langle v_\theta^2
  \right \rangle$ provides a direct measure for the typical velocities
  of convective and SASI motions and the violence of these
  instabilities.
\label{fig:convective_energy}
}
\end{figure*}

\begin{figure}
\plotone{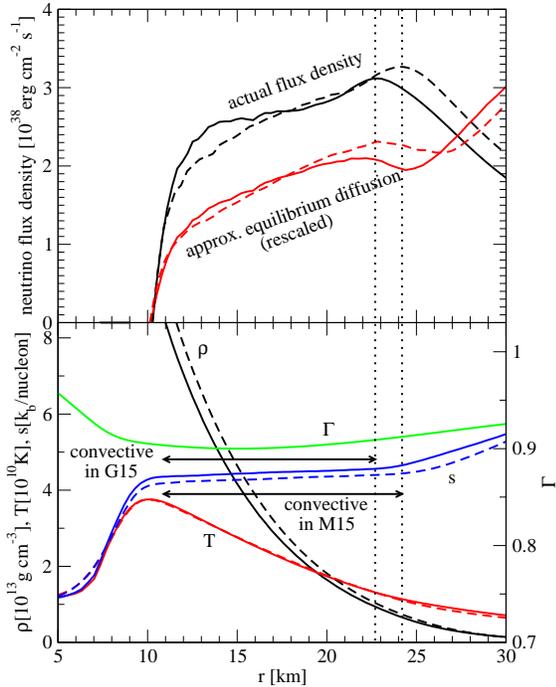}
\caption{Origin of the different $\nu_\mu$ and $\nu_\tau$ luminosities
  in models G15 and M15. The upper panel shows the energy flux density
  of $\nu_\mu$ and $\nu_\tau$ at $250 \ \mathrm{ms}$ after bounce
  (black lines) for G15 (solid) and M15 (dashed) and illustrates that
  a smaller maximum flux is reached at a somewhat smaller
  ``freeze-out'' or ``number sphere'' radius in model G15
  (corresponding to a saturation of the luminosity at a lower value).
  Dotted lines are used to denote the rough location of the freeze-out
  radius in both models. The lower flux at radii around $24
  \ \mathrm{km}$ is consistent with an estimate of the flux
  $H_{\nu,\mathrm{diff}}$ expected for equilibrium diffusion
  (Equation~\ref{eq:diffusive_flux}) as indicated by the red curves:
  While the diffusive fluxes agree well within a large part of the
  convection zone, the shift of the ``dip'' in the predicted
  $H_{\nu,\mathrm{diff}}$ in G15 suggests a lower freeze-out radius
  and luminosity. Note that the diffusive flux has been rescaled to
  avoid overlap with the black curve. Since the density- and
  temperature-dependence of the diffusion coefficient cannot be
  captured exactly by simple power laws, a perfect fit is not to be
  expected anyway; Equation~(\ref{eq:diffusive_flux}) is rather used
  to detect \emph{relative} differences between the two models G15 and
  M15. Angle-averaged profiles of the quantities used to evaluate
  Equation~(\ref{eq:diffusive_flux}), i.e.\ the density $\rho$, the
  temperature $T$, and the metric factor $\Gamma$ are shown in the
  lower panel along with the angle-averaged entropy $s$. The different
  fluxes in models G15 and M15 stem from several (partially competing)
  factors, namely a shallower temperature gradient in model G15 from
  $22 \ \mathrm{km}$ outward, which is associated with an earlier rise
  of $s$ outside the proto-neutron star convection zone, lower
  densities (which increase the diffusive flux; the relative
  difference is still $\sim 10\%$ at $25 \ \mathrm{km}$), and the
  reduction of the flux due the $\Gamma$-factor (which is not present
  in M15). The lower freeze-out radius in model G15 is associated with
  a different location of the edge of the (roughly adiabatically
  stratified) PNS convection zone, which is clearly visible in the
  entropy profile.
\label{fig:mu_tau_luminosity}
}
\end{figure}

\subsection{$15 M_\odot$ Progenitor -- Influence of General Relativity}
\label{sec:gr_effects}

The case of the $15 M_\odot$ model is much more interesting because
the different outcome of model G15 as opposed to M15, N15, and S15
suggests that this progenitor is a marginal case and may reveal the
dependence of the explosion conditions on the input physics used in
these three runs, i.e.\ the influence of the GR treatment and the
neutrino reaction rates. Indeed, a comparison of the runaway criterion
$\tau_\mathrm{adv}/\tau_\mathrm{heat}$ and the critical time-scales
(Figure~\ref{fig:timescale_comparison}) shows systematic differences:
For model G15, we consistently find higher values of
$\tau_\mathrm{adv} / \tau_\mathrm{heat}$ than in the other $15
M_\odot$ runs, and this model is also characterized by a larger
heating efficiency $\dot{Q}_\mathrm{heat}/(L_{\nu_e}+L_{\bar{\nu}_e})$
(Figure~\ref{fig:heating_efficiency}), i.e.\ a larger fraction of energy
radiated in $\nu_e$ and $\bar{\nu}_e$ is re-absorbed in the gain
layer.\footnote{Note that in contrast to \citet{marek_09},
we use the luminosities of $\nu_e$ and $\bar{\nu}_e$ at the gain
radius (see Figure~\ref{fig:neutrino_gain_comparison}) instead
of the (lower) values for an observer at infinity.} Moreover, the volume-integrated heating rate 
  $\dot{Q}_\mathrm{heat}$ is also highest among the $15 M_\odot$ runs
(Figure~\ref{fig:gain_region_comparison}).

Compared to model M15, $\tau_\mathrm{adv} / \tau_\mathrm{heat}$ is
typically higher by a factor of two, resulting from the combination of
both a longer advection time-scale and a shorter heating
time-scale. The purely Newtonian model N15 exhibits a very similar
evolution of the runaway criterion as model M15, but this is due to a
cancellation of huge differences (amounting to a factor of $2 \ldots
3$) in both the advection and the heating time-scale. The relativistic
calculation shows, however, that even though the effects of a smaller
shock radius (Figure~\ref{fig:s15_shock_radius}) and hence a shorter
advection time-scale on the one hand and stronger heating on the other
hand \emph{partially} compensate each other in GR, such a cancellation
is not to be taken for granted, and the residual effect, e.g.\ on the
time-scale ratio $\tau_\mathrm{adv}/\tau_\mathrm{heat}$ may still be
on the order of several tens of percents.

For model S15, the time-scale ratio lies about half way in between G15
and N15, suggesting a non-negligible effect of the neutrino
microphysics. Considering the claims of \citet{bruenn_09} to that
effect, this is certainly noteworthy, but we shall first turn our
attention to the more pronounced differences between the GR model G15
and the runs M15 and N15 with a different treatment of gravity.
Since our focus lies on the GR effects in this section, we
defer the discussion of model S15 to Section~\ref{sec:neutrino_effects},
however.

\subsubsection{Newtonian Approximation vs. General Relativity}

The purely Newtonian case (N15) stands apart most clearly from the
others with large values of $\tau_\mathrm{adv}$ and
$\tau_\mathrm{heat}$, and the reason for this has essentially been
given by \citet{bruenn_01}: Due to the shallower potential, the
proto-neutron star is considerably more extended in model N15 compared
to the GR case (Figure~\ref{fig:pns_radius_and_temperature}), and this
also shifts the gain radius and shock radius further out
(Figures~\ref{fig:s15_shock_radius} and
\ref{fig:gain_region_comparison}), thus increasing the mass
  $M_\mathrm{gain}$ in the gain layer. On the other hand, the neutron
star surface temperature is also considerably lower compared to GR
(Figure~\ref{fig:pns_radius_and_temperature}), resulting in a
significant reduction of the neutrino luminosities and mean energies
in the gain region (Figure~\ref{fig:neutrino_gain_comparison}) and hence
weaker heating in the gain region
(Figure~\ref{fig:gain_region_comparison}). Apparently this also leads to less
vigorous convection in the purely Newtonian case as the mass-specific kinetic energy
contained in non-radial mass motions, $E_\mathrm{kin,\theta}/M_\mathrm{gain}$, is
typically lower than in model G15
(Figure~\ref{fig:convective_energy}) by $40\% - 50\%$.  Stronger
convection in GR partly compensates for the reduction of
$M_\mathrm{gain}$ and $\tau_\mathrm{adv}$ due to the smaller neutron
star radius and therefore helps to turn the scales in favor of a
larger value of the runaway criterion
$\tau_\mathrm{adv}/\tau_\mathrm{heat}$ -- the large reduction of
$\tau_\mathrm{heat}$ emerges as the dominant effect (Figure~\ref{fig:timescale_comparison}).

In comparing model N15 to G15 and M15, we should also bear in mind
that N15 was computed with a higher angular resolution of 128 zones,
which might be beneficial for the heating conditions because it was
seen to foster explosions in 2D simulations of \citet{hanke_11}. The
increase of $\tau_\mathrm{adv}/\tau_\mathrm{heat}$ in GR compared to
the Newtonian approximation may therefore even be underestimated by
our analysis.

\newpage
\subsubsection{Effective Potential Approximation vs. General Relativity}
\label{sec:eff_potential}

The comparison between models G15 (GR) and M15 (effective potential)
is somewhat more subtle, but the different heating conditions can
still be traced back -- at least partly -- to the neutrino emission
from the proto-neutron star. Again, the neutrino luminosities and mean
energies at the gain radius (Figure~\ref{fig:neutrino_gain_comparison})
turn out to be the crucial factor.  In the early phase, the GR run
exhibits a noticeable enhancement in the electron antineutrino
luminosity (by $\approx 15 \%$) and to a lesser extent in the electron
neutrino luminosity. In addition, the mean energies of
$\nu_\mathrm{e}$ and $\bar{\nu}_\mathrm{e}$ tend to increase more
strongly at late times, with the difference reaching almost $1
\ \mathrm{MeV}$ for the antineutrinos.

Interestingly, the tendency towards slightly more energetic
$\nu_\mathrm{e}$'s and $\bar{\nu}_\mathrm{e}$'s in GR is already
present in 1D (see Paper I). This is presumably the result of a
slightly different density stratification in GR that cannot be
reproduced exactly by the modified Newtonian potential {and the
  approximate GR transport treatment (e.g.\ due to the identification
  of coordinate radius and proper radius) used for the M15 run
  (cp.\ \citealp{marek_06}). The circumferential radius of the PNS
  (defined as the radius where the density drops to $10^{11}
  \ \mathrm{g} \ \mathrm{cm}^{-3}$) is indeed larger by 2\%--4\% in
  GR, and its surface is somehwat hotter at late times
  (Figure~\ref{fig:pns_radius_and_temperature}). In contrast to the
  higher $\nu_e$ and $\bar{\nu}_e$ luminosities in model G15, the
  luminosity of $\mu$ and $\tau$ neutrinos is smaller in the GR
  case. This is the result of general relativistic transport
    effects and a different stratification at the edge of the PNS
  convection zone, where the ``number sphere'' for $\nu_\mu$ and
  $\nu_\tau$ (i.e.\ the sphere up to which production processes and
  absorption processes remain in equilibrium; see
  \citealp{raffelt_01,keil_03}) is located, as we illustrate for a
  representative snapshot $250 \ \mathrm{ms}$ after bounce
  (Figure~\ref{fig:mu_tau_luminosity}). The upper panel of
  Figure~\ref{fig:mu_tau_luminosity} shows the energy flux density
  $H_\nu$ of $\nu_{\mu/\tau}$ in the comoving frame for models G15 and
  M15 at that time, with $H_\nu$ rising to somewhat higher values
  before it decays like $r^{-2}$ outside the ``number
  sphere''.\footnote{ We remark that the luminosity emerging from the
    ``number sphere'' is smaller than the black body luminosity by one
    to two orders of magnitude due to a very small flux factor (see
    discussion in \citealt{janka_95b}). Stefan's law is therefore
    ill-suited for estimating the luminosity of $\nu_\mu$ and
    $\nu_\tau$.} To demonstrate that the smaller neutrino flux in the
  case of model G15 is consistent with the temperature and density
  stratification depicted in the lower panel of
  Figure~\ref{fig:mu_tau_luminosity} we also show a rough estimate for
  the diffusive flux for comparison in the upper panel. The diffusive
  flux $H_{\nu,\mathrm{diff}}$ can be computed as (see, e.g.,
  \citealp{pons_99}, Equation 13)\footnote{In \textsc{Vertex-CoCoNuT}, we
    do not distinguish between heavy flavor neutrinos and
    antineutrinos, and the chemical potential $\mu_{\nu_{\mu/\tau}}$
    therefore vanishes.  Consequently, only derivatives of the
    temperature appear in the diffusion equation.}
\begin{equation}
\label{eq:diffusive_flux}
H_{\nu,\mathrm{diff}}=k D \frac{T^3}{\alpha} \left(\Gamma \frac{\partial \alpha T}{\partial r}\right)
\end{equation}
using Schwarzschild radial coordinates (with $r$ denoting the
circumferential radius). Here $\alpha$ is the lapse function, $D$ an
appropriate, frequency-averaged diffusion coefficient, and some
factors of order unity have been absorbed into the constant $k$ for
convenience. A metric factor $\Gamma$ appears because the diffusive
flux depends on the derivative $\Gamma \partial/\partial r$ with respect to
\emph{proper} (physical) radius.\footnote{In \textsc{Vertex-CoCoNuT}
  we work with a different gauge choice for the radial coordinate
  (isotropic coordinates $r_\mathrm{iso}$), but the conversion to the
  familiar Schwarzschild form of the metric is straightforward.
  $\Gamma$ can be obtained from the conformal factor $\phi$ as
  $\Gamma=1+2 r_\mathrm{iso} \partial \ln \phi/\partial r_\mathrm{iso} $.}  In
model M15, no distinction is made between proper radius and
circumferential radius ($\Gamma=1$). As scattering on nucleons is the
dominant source of opacity, the diffusion coefficient $D$ is roughly
proportional to $\rho^{-1} T^{-2}$ in the density regime that we are
considering here. Figure~\ref{fig:mu_tau_luminosity} demonstrates that
Eq.~(\ref{eq:diffusive_flux}) predicts different fluxes in the region
around the ``number sphere'' and also points to a slightly smaller
``number sphere'' radius in G15, which is associated with a different
location of the outer boundary of the PNS convection zone
(Figure~\ref{fig:mu_tau_luminosity}). The lower flux, mainly due to
the factor $\Gamma$ in the GR case and the smaller ``number sphere''
radius account for the $\sim 20\%$ lower $\nu_\mu$ and $\nu_\tau$
luminosity in model G15. On the other hand, there is no such reduction
of the neutrino mean energy of $\nu_\mu$ and $\nu_\tau$ in model G15,
because neutrino-electron scattering and non-isoenergetic
neutrino-nucleon scattering still allow for energy exchange and
thermal equilibration with the medium at larger radii, where the
temperature is \emph{higher} than in model M15.

While GR transport and stratification effects thus seem to account for
the different neutrino emission in models G15 and M15 and, because of
the higher $\nu_e$ and $\bar{\nu}_e$ luminosities, suggest somewhat
better heating conditions, the huge increase of the time-scale
criterion in model G15 still needs to be connected to these
findings. Merely reducing the heating time-scale by $\approx 20 \%$ in
model G15 compared to M15 would not be sufficient to ensure that the
runaway condition $\tau_\mathrm{adv} / \tau_\mathrm{heat}>1$ is met.
At $450 \ \mathrm{ms}$ the crucial factor distinguishing the
relativistic and the pseudo-Newtonian run is the larger value of
$M_\mathrm{gain}$ (Figure~\ref{fig:gain_region_comparison}) and the
longer advection time-scale
$\tau_\mathrm{adv}=M_\mathrm{gain}/\dot{M}$
(Figure~\ref{fig:timescale_comparison}).  Part of this increase in
$\tau_\mathrm{adv}$ is a direct consequence of the higher neutrino
luminosities, as the increase in thermal pressure behind the shock
allows for a larger mass in the gain region.  Since the position of
the gain radius is almost identical in model G15 and M15
(Figure~\ref{fig:gain_region_comparison}), this can be illustrated
qualitatively on the basis of a very rough approximation: Requiring
balance between heating and cooling at the gain radius with heating
and cooling rates per nucleon roughly proportional to $L \left\langle
E_\nu^2\right\rangle /r^2$ and $T^6$, respectively
(e.g.\ \citealp{bethe_85,janka_01}), and assuming that the gas at the
gain radius is radiation-dominated, we find that the pressure
$P_\mathrm{gain}$ and the temperature $T_\mathrm{gain}$ at the gain
radius scale as
\begin{equation}
P_\mathrm{gain}^{3/2}\propto T_\mathrm{gain}^6 \propto \frac{L_\nu
  \left\langle E_\nu^2\right\rangle}{r_\mathrm{gain}^2}.
\end{equation}
The stratification in the gain region is roughly adiabatic with the
pressure and density following power laws ($P \propto r^{-4}$, $\rho
\propto r^{-3}$), and in spherical symmetry we can therefore
approximatively determine the radius of the stagnant accretion shock
with the help of the jump conditions at the shock.  For the post-shock
pressure $P_\mathrm{post}$, we have
\begin{equation}
P_\mathrm{post}=\left(1-\frac{1}{\beta}\right) P_\mathrm{ram},
\end{equation}
where $P_\mathrm{ram}$ is the ram pressure ahead of the shock, and
$\beta$ is the ratio of the post- and the pre-shock
density. $P_\mathrm{ram}=\rho_\mathrm{pre} v_\mathrm{pre}^2$ can be
computed assuming that the pre-shock velocity $v_\mathrm{pre}$ is a
certain fraction of the free-fall velocity, i.e.\ $v_\mathrm{pre}
\propto 1/\sqrt{r_\mathrm{sh}}$, and therefore scales as
\begin{equation}
P_\mathrm{ram} = \frac{\dot{M} v_\mathrm{pre}}{4 \pi r_\mathrm{sh}^2}\propto \frac{\dot{M}}{r_\mathrm{sh}^{5/2}}.
\end{equation}
On the other hand, the post-shock pressure is related to the
pressure at the gain radius roughly by 
\begin{equation}
P_\mathrm{post} r_\mathrm{sh}^4 \approx P_\mathrm{gain} r_\mathrm{gain}^4.
\end{equation}
Assuming $r_\mathrm{gain}$ to be fixed (motivated by
Figure~\ref{fig:gain_region_comparison}), the shock radius therefore
varies with $L_\nu \left\langle E_\nu^2 \right\rangle$ as
\begin{equation}
\label{eq:r_shock_scaling}
r_\mathrm{sh} \propto \left(L_\nu \left\langle E_\nu^2 \right\rangle \right)^{4/9}.
\end{equation}
Using Equation~(\ref{eq:r_shock_scaling}), we can compute the mass
$M_\mathrm{gain}$ in the gain region for a density stratification with
$\rho \propto r^{-3}$,
\begin{equation}
M_\mathrm{gain} \propto \int\limits_{r_\mathrm{gain}}^{r_\mathrm{sh}} \frac{\beta \dot M}{v_\mathrm{pre} r_\mathrm{sh}^2}
\left(\frac{r_\mathrm{sh}}{r}\right)^3 r^2 \,\mathrm{d} r
\propto
\dot{M} r_\mathrm{sh}^{3/2} \ln\left(\frac{r_\mathrm{sh}}{r_\mathrm{gain}}\right).
\end{equation}
The logarithmic derivative,
\begin{equation}
\label{eq:sensitivity_m_gain}
\frac{\partial \ln M_\mathrm{gain}}{\partial \ln \left(L_\nu \left\langle E_\nu^2 \right\rangle \right)}
\approx
\frac{2}{3}+\frac{4}{9 \ln (r_\mathrm{sh}/r_\mathrm{gain})}
\sim 1,
\end{equation}
can be taken as a measure for the sensitivity of $M_\mathrm{gain}$ to
changes in $L_\nu \left\langle E_\nu^2 \right\rangle$. Although based
on a rather crude approximation, Equation~(\ref{eq:sensitivity_m_gain})
suggests that changes in the luminosity and neutrino energy induce
comparably large changes in $M_\mathrm{gain}$ and hence
$\tau_\mathrm{adv}$.

In addition, stronger neutrino heating also leads to more violent
activity of non-radial mass motions in the form of convection and the
SASI (see Figure~\ref{fig:convective_energy}). This in turn increases
the residence time-scale of matter in the gain region and thus further
boosts the effect of the stronger neutrino heating in model G15. With
both mechanisms working in combination, the mass in the gain region
$M_\mathrm{gain}$ and hence the advection time-scale
$\tau_\mathrm{adv}$ reach considerably larger values in model G15
(except for a short quiet period (see
Figures~\ref{fig:entropy_poles},\ref{fig:s15_shock_radius}) with little
SASI activity between $350 \ \mathrm{ms}$ and $400
\ \mathrm{ms}$). Likewise, the energy deposition rate in the gain
region (Figure~\ref{fig:gain_region_comparison}) remains higher than in
model M15 by a factor of $\approx 2$ from $200 \ \mathrm{ms}$ onward.

While we view all these aspects as a likely explanation, it
cannot be excluded that other factors are responsible for the
different evolution of the shock after the accretion of the Si/SiO and
the relatively quiet period around $350 \ \mathrm{ms}$. Different
growth and saturation properties of the SASI due to general
relativistic effects could be invoked as an alternative explanation,
although this seems rather far-fetched considering that the non-linear
behavior of the SASI is still under investigation even in the
Newtonian case (where parasitic {Kelvin-Helmholtz or Rayleigh-Taylor}
instabilities have been proposed as a saturation mechanism by
\citealt{guilet_10}). The possibility of numerical effects (in a broad
sense) ought to be considered more seriously: The hydro solvers
\textsc{CoCoNuT} and \textsc{Prometheus} used for model G15 and M15,
respectively (Table~\ref{tab:model_setup}), employ different Riemann
solvers, \textsc{Prometheus} relies on dimensional splitting, the
radial grids are not identical, and neither are the initial
perturbations. However, there is little grounds for relating these
technical differences to the behavior of models G15 and M15. Apart
from the fact that the Si/SiO interface is preserved a little more
sharply in \textsc{CoCoNuT} (which may account for a slightly stronger
reaction of the shock to the reduced accretion rate,
Figure~\ref{fig:s15_shock_radius}), we have no tangible evidence for
such a connection. Moreover, the fact that the\ \textsc{CoCoNuT} model
S15 -- with a less pessimistic ratio
$\tau_\mathrm{adv}/\tau_\mathrm{heat}$
(Figure~\ref{fig:timescale_comparison}) -- also responds less vigorously
to the drop in the accretion rate than model G15 points to a physical
origin as well. It therefore seems that the more rigorous GR treatment
is ultimately the reason for the more optimistic evolution of model
G15 compared to the pseudo-Newtonian simulation M15.

\begin{figure*}
\plotone{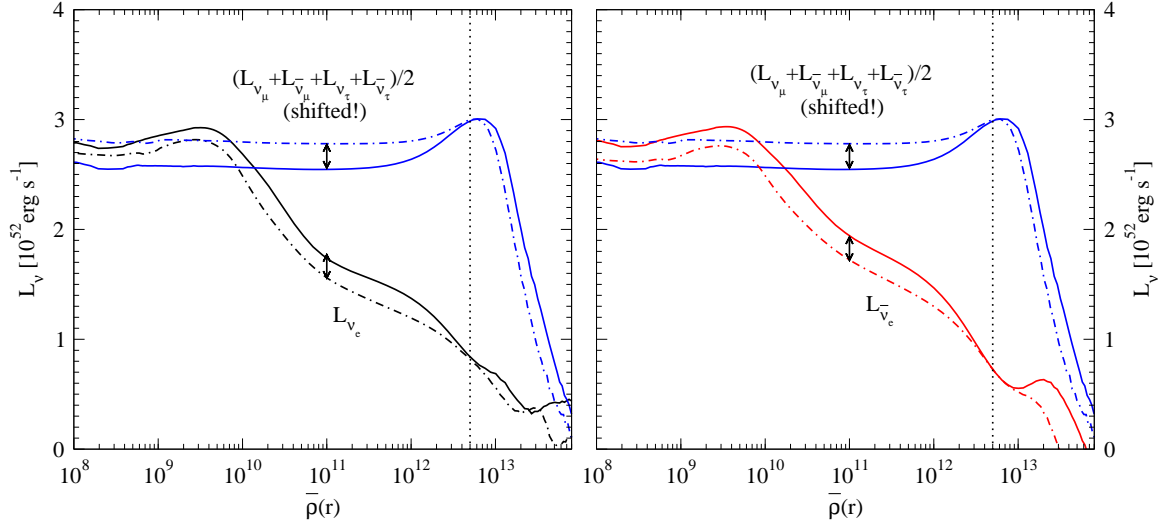}
\caption{Illustration of enhanced $\nu_e$ and $\bar{\nu}_e$ emission
  as an indirect consequence of non-isoenergetic nucleon recoil
  effects in the $\nu_{\mu/\tau}$ sector. The two panels show the
  redshifted lab-frame luminosities of $\nu_e$ (left panel, black
  curves) and $\bar{\nu}_e$ (right panel, red curves) at a time of
  $400 \ \mathrm{ms}$ after bounce as a function of the average
  density $\bar{\rho}(r)$ at a given radius $r$ for model G15 (black solid
  lines, non-isoenergetic nucleon recoil included) and model S15
  (dash-dotted lines, without energy losses of neutrinos in nucleon
  scatteringsq). At $\bar{\rho}(r) \approx 5 \times 10^{12}
  \ \mathrm{g} \ \mathrm{cm}^{-3}$ (indicated by the dotted vertical
  line), the $\nu_e$ (and, respectively, the $\bar{\nu}_e$)
  luminosities with and without nucleon recoil are roughly identical.
  For the heavy-flavor neutrinos (${\nu}_\mu$, $\bar{\nu}_\mu$,
  ${\nu}_\tau$, and $\bar{\nu}_\tau$), we show \emph{half the total
    luminosity} of the four species and \emph{shift} the resulting
  curves to a value of $3 \times 10 ^{52} \ \mathrm{erg}
  \ \mathrm{s}^{-1}$ at $\bar{\rho}(r) \approx 5 \times 10^{12}
  \ \mathrm{g} \ \mathrm{cm}^{-3}$ (blue curves) in order to
  illustrate the different amount of energy transfer from
  $\nu_{\mu/\tau}$ to the medium below this density. Energy transfer
  from ${\nu}_{\mu/\tau}$ (reflected by the decrease of $L_\nu$
  towards lower densities) to the background medium occurs roughly
  between $5 \times 10^{12} \ \mathrm{g} \ \mathrm{cm}^{-3}$ and a few
  $10^{11} \ \mathrm{g} \ \mathrm{cm}^{-3}$. With non-isoenergetic
  nucleon recoil, the energy input into the medium is larger, and
  there is a corresponding enhancement of the $\nu_e$ and
  $\bar{\nu}_e$ luminosity in precisely the same region. If the
  additional energy gained from ${\nu}_{\mu/\tau}$ scattering were
  completely transferred into and equally split between ${\nu}_e$ and
  $\bar{\nu}_e$, the expected enhancement of the luminosity would be
  $2 \times 10^{51} \ \mathrm{erg} \ \mathrm{s}^{-1}$ (corresponding
  to the length of the double arrows). This is in reasonable agreement
  with the observed values.
\label{fig:redistribution}
}
\end{figure*}

\subsubsection{Differences to Previous Pseudo-Newtonian Simulations}
\citet{marek_09} already investigated the $15 M_\odot$ progenitor of
\citet{woosley_95} in the framework of the effective potential
approximation using the same ray-by-ray variable Eddington factor
method that has been applied in our study. Given the close similarity
of their model setup, the simulations presented here must also be
interpreted against the background of their earlier results.

\citet{marek_09} considered four different simulations of the $15
M_\odot$ progenitor, namely two rotating models (128 angular zones)
with different choices for the effective gravitational potential and
different resolution in energy space, and two highly-resolved runs
(192 angular zones) without rotation and with two different equations
of state. Only one of these, the model including rotation and a
gravitational potential that somewhat overestimates strong-field
effects (case~R of \citealt{marek_06}), developed an explosion roughly
$550 \ \mathrm{ms}$ after bounce.

By combining our findings with those of \citet{marek_09}, we arrive at
a rather coherent picture. In both cases, the $15 M_\odot$ progenitor
appears to evolve very close to the explosion threshold, as small
model variations were sufficient to bring about an explosion, while
other simulations with slightly less favorable showed no signal
of shock revival at least until $\gtrsim 400 \ \mathrm{ms}$ after
bounce.  The qualitative similarities between our results and those of
\citet{marek_09} go much further: It is remarkable that in both
  cases the explosion is not linked to the transition of the Si/SiO
  interface through the shock but occurs at a much later stage when
  the SASI again becomes violent after an intermediate phase of
  relatively weak activity. In both cases, the most optimistic model
is one with stronger neutrino emission than for the ``best'' effective
potential (case~A) of \citet{marek_06}, which is either due to the
better treatment of GR (for our model G15) or the choice of a stronger
effective potential (case~R) for the explosion model LS-rot of
\citet{marek_09}\footnote{The explosion model LS-rot of
  \citet{marek_09} also includes rotation, which, as these authors
  found, adversely affects the heating conditions. This mitigates the
  effect of the overly strong gravitational potential and makes their
  model explode at a similarly late stage as model G15, although with
  an extra delay of $\gtrsim 100 \ \mathrm{ms}$.}.  In the light of
their earlier results, the beneficial role of GR corrections thus
seems all the more plausible.

The work of \citet{marek_09} also provides us with reference runs
computed with a higher angular resolution, and thus allow us to
address an important limitation of the present first generation of GR
neutrino hydrodynamics models. In particular, they discuss a
non-rotating effective potential model (LS-2D) corresponding exactly
to model M15, but with a higher resolution of 192 angular zones (as
compared to 64). For the first $200 \ \mathrm{ms}$ after bounce, the
shock evolution, as well as the advection and heating time-scales
agree remarkably well, while the high-resolution run of
\citet{marek_09} shows a somewhat more optimistic evolution
afterwards. This suggests that our findings about the beneficial
effects of GR will prove robust for two reasons: Differences between
the GR model G15 and the effective potential run M15 apparently assert
themselves earlier than resolution effects and also appear to be more
pronounced. Moreover, the comparison of model M15 and model LS-2D of
\citet{marek_09} provides further evidence for the more optimistic
evolution of high-angular resolution runs in 2D that has been found by
\citet{hanke_11}. It therefore seems likely that high-resolution
follow up studies will confirm our present results about explosions in
GR.

\subsection{Influence of the Neutrino Interaction Rates}
\label{sec:neutrino_effects}
Besides general relativity, the treatment of the neutrino-matter
interactions is another potential factor that can influence the
heating conditions and the evolution towards an explosion. Models S15
and G15 provide a comparison between an older ``simplified'' set of
opacities and an up-to-date treatment. These models
serve to illustrate the
importance of the neutrino microphysics for the dynamics, thus
contributing to an ongoing debate about the necessary level of
sophistication in the neutrino treatment \citep{nordhaus_10,lentz_12a,lentz_12b}.
In the following, we present an analysis of the most conspicuous
differences between the ``simplified'' and ``full'' rates with respect
to the dynamics of the accretion phase. Due to the computational
cost of multi-dimensional neutrino hydrodynamics simulations,
we cannot attempt a systematic study of all individual rates; such
a systematic investigation is presently only feasible in 1D, where
it has recently been carried out by \citet{lentz_12b}.

Both the shock trajectory (Figure~\ref{fig:s15_shock_radius}) and the
time-scale ratio $\tau_\mathrm{adv}/ \tau_\mathrm{heat}$
(Figure~\ref{fig:timescale_comparison}) clearly indicate that model
G15 with the full set of rates evolves more optimistically than model
S15 computed with the simplified rates.  Changing the whole
``package'' of neutrino interaction rates thus appears to influence
the dynamical evolution during the accretion phase quite noticeably.
As in Section~\ref{sec:gr_effects}, the neutrino emission from the
proto-neutron star (Figure~\ref{fig:neutrino_gain_comparison})
provides the clue for understanding these differences.

The most conspicuous effect
in model G15 is the reduction of the $\nu_\mathrm{\mu/\tau}$ mean
energies (because of the additional inclusion of non-isoenergetic neutrino-nucleon scatterings,
see below). This effect is of little direct relevance for the energy
deposition in the gain layer to which $\nu_\mu$ and $\nu_\tau$ hardly
contribute. However, electron neutrinos and antineutrinos are also
affected on a smaller scale with some repercussions on the heating in
the gain layer. In particular, electron antineutrinos are emitted with
higher luminosity (by about $10\%$) and -- towards the later phases --
higher mean energy (by up to $0.8\ \mathrm{MeV}$) in model
  G15. For the electron neutrinos, the luminosity enhancement is
smaller, and the spectra are only a little harder with the improved
set of interaction rates.

However, Figure~\ref{fig:timescale_comparison} shows that using the
improved opacities results only in a relatively insignificant decrease
of the heating time-scale in G15. This small effect is magnified by
the increase of the advection time-scale due to a larger average shock
radius (see Figure~\ref{fig:s15_shock_radius} and the discussion in
Section~\ref{sec:eff_potential}), as well as by increased convective and
SASI activity (see Figure~\ref{fig:convective_energy}), and therefore
still leads to an appreciable difference in the time-scale ratio
$\tau_\mathrm{adv}/\tau_\mathrm{heat}$ between models G15 and S15.
The reduced heating in model S15 thus at least results in a
considerable delay of a possible explosion in this model.

It remains to be discussed how the improved interaction rates in model
G15 contribute to an enhancement of $\nu_e$ and $\bar{\nu}_e$
emission.  G15 and S15 differ in the treatment of a number of
interaction processes (see~Table~\ref{tab:rates}). Based on two
models, we can naturally pin down only the dominant effect relevant
for the change of the heating conditions. The reader should note,
however, that all of the individual processes for which we use an
improved treatment in model G15 (and not only these) are relevant in
their own rights in other contexts, as has already been documented in
the literature: The electron capture rates on heavy nuclei of
\citet{langanke_03}, e.g., lead to stronger deleptonization during
collapse, reduce the mass of the homologous core, and result in the
formation of a slightly weaker shock at bounce \citep{hix_03}. The
inclusion of nucleon correlations strongly decreases the opacity at
high densities \citep{burrows_98,burrows_99,pons_98,pons_99} and thus
shortens the proto-neutron star cooling time considerably
\citep{huedepohl_10}.  Neutrino pair conversion can somewhat enhance
the emission of $\nu_\mu$ and $\nu_\tau$ from the proto-neutron star
\citep{buras_03}. We refer the reader to the literature for a more
detailed discussion of the influence of the individual interaction
processes on the neutrino emission and their impact on the dynamics in
1D. \citet{lentz_12b}, in particular, recently studied 
the interplay of rate variations in great detail in 1D and gave
a succinct summary of possible rate effects.

If the dynamical differences between models G15 and S15 are to be
explained by neutrino rate effects, we need a process that can
significantly change -- perhaps indirectly -- the electron neutrino
and antineutrino emission during the accretion phase.  As a closer
analysis shows, among the improvements of interaction rates listed in
Table~\ref{tab:rates} (which include all the aforementioned
processes), only energy transfer from $\mu$ and $\tau$ neutrinos to
nucleons by scattering reactions in the PNS surface region (not in the
gain region!) can achieve such a change and emerges as the most likely
cause of the dynamical differences between S15 and G15. As pointed
out by several authors \citep{janka_89,suzuki_90,raffelt_01,keil_03},
$\nu_\mu$ and $\nu_\tau$ pass an extended scattering atmosphere
(dominated by elastic neutrino-nucleon scattering reactions), which
separates the ``number sphere'' where the production reactions (in our
case mainly bremsstrahlung and neutrino-neutrino pair conversion)
freeze out, and the ``transport sphere'' which marks the transition to
free streaming. In this intermediate region, mainly nucleon recoil in
scattering reactions still allows for a certain amount of energy
exchange between the neutrinos and the medium, which reduces the mean
energies of $\nu_\mu$ and $\nu_\tau$ appreciably
(Figure~\ref{fig:neutrino_gain_comparison}).  The
  enhancement of the neutrino luminosity due to the inclusion of
  neutrino pair conversion \citep{buras_03} is also largely canceled
  by this energy exchange (although the \emph{number} flux of
  $\nu_\mu$ and $\nu_\tau$ is still considerably larger in model G15
  compared to S15, see also \citealt{keil_03}).

Our simulations show that ``downscattering'' effect on the spectra of
$\nu_{\mu/\tau}$ has further consequences if the energy exchange
between the neutrinos and the background medium is included
self-consistently.  Neutrino absorption and emission in the PNS
surface region maintain a quasi-steady-state temperature
stratification which is determined by the neutrino flux temperature
and thus by the temperature of regions deeper in the neutron star
(which is similar in models G15 and S15). In order to maintain this
stratification, any additional energy input from $\nu_{\mu/\tau}$ at
densities around $10^{12} \ \mathrm{g} \ \mathrm{cm}^{-3}$ must be
compensated by a corresponding energy loss in $\nu_e$ and
$\bar{\nu}_e$.  As this rather efficient conversion of $\nu_{\mu/\tau}$
``recoil energy'' of the nucleons into $\nu_e$ and $\bar{\nu}_e$
happens in a relatively hot layer, there is also a (minor) enhancement
of the mean energies of $\nu_e$ and $\bar{\nu}_e$.  We illustrate this
``reshuffling'' of energy between the different flavors in
Figure~\ref{fig:redistribution}, which shows that the luminosity
decrease in $\nu_{\mu/\tau}$ due the scattering losses to nucleons and
the enhancement of the $\nu_e$ and $\bar{\nu}_e$ luminosity occur in
precisely the same density region and correspond well in magnitude.

It should be emphasized that the additional energy transfer to the
medium in the nucleon scattering reactions is crucial for this effect, which cannot be accounted for by
the mere change of the scattering and absorption opacity due to the
reduction of the final lepton phase space \citep{schinder_90,
  horowitz_02}. We also point out that the inclusion of the nucleon
recoil in the charged-current processes has little effect on the
heating by $\nu_e$ and $\bar{\nu}_e$: Here, the complete energy of
the absorbed neutrino is transferred to the medium anyway -- a
different partitioning of the deposited energy between nucleons and
electrons/positrons does not increase the total energy transfer to the
medium. The only change comes from the reduction of the total
cross-section \citep{horowitz_02}, which is small at the relevant
neutrino energies between roughly $10 \ \mathrm{MeV}$ and $20
\ \mathrm{MeV}$.

All in all, our analysis highlights the need to attend to the
details for neutrino-matter interactions in order to model
the dynamics of the pre-explosion phase with high accuracy,
demonstrating that even some seemingly minor corrections in
the rates can have a non-negligible impact on the
heating conditions.

\subsection{An Aside on Other Measures for the Criticality of the Accretion Flow}
\begin{figure}
\plotone{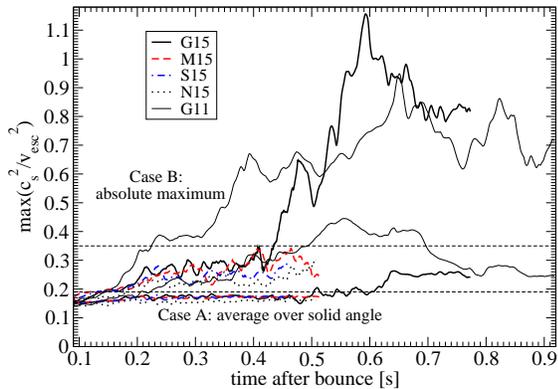}
\caption{Time evolution of the maximum value of the ratio of the sound
  speed $c_s$ and the (local) escape velocity $v_\mathrm{esc}$ for
  models G11, G15, M15, S15, and N15, smoothed over $10
  \ \mathrm{ms}$. Two different methods for evaluating $\max
  (c_s^2/v_\mathrm{esc}^2)$ are used: We either compute spherical
  averages of $c_s^2$ before taking the maximum (case~A) or take the
  global maximum (case~B). Only the region outside the PNS
  ($\rho<10^{11} \ \mathrm{g} \ \mathrm{cm}^{-3}$) is considered in
  either case. The escape velocity is computed directly from the
  gravitational potential or, in the GR case, the lapse function.
\label{fig:pt_criterion}
}
\end{figure}

Extending earlier work on the subject \citep{burrows_93,yamasaki_07},
\citet{pejcha_11} recently re-investigated the problem of a stationary
1D accretion flow onto a proto-neutron star in great detail, and
attempted to derive firm criteria for the transition from the phase of
quasi-stationary accretion to the explosion. They claimed to have
found such a precise and robust criterion in the so-called antesonic
condition for the maximum ratio of the local sound speed $c_s$ and the
escape velocity $v_\mathrm{esc}$ in the post-shock region, stating
\begin{equation}
\max{ \left(\frac{c_s^2}{v_\mathrm{esc}^2}\right)  > \frac{3}{16}}
\end{equation}
as requirement for an explosion. According to their results, this
antesonic condition is a more reliable measure for the criticality of
the flow than other criteria such as the ratio
$\tau_\mathrm{adv}/\tau_\mathrm{heat}$ used in
Section~\ref{sec:heating}, or the requirement of a growing mass and
energy in the gain region \citep{janka_01}. Very recently,
\citet{fernandez_12} gave a very careful assessment of the
Burrows-Goshy-limit and its relation to the conditions where a
steady-state accretion flow in 1D meets the threshold
for a runaway instability.

Considering the ongoing debate about explosion criteria, it is
worthwhile to study the applicability of the new antesonic condition
to dynamical multi-dimensional simulations of the accretion phase, for
which the assumptions of \citet{pejcha_11} (e.g.\ spherical symmetry,
stationarity) do not necessarily constitute a valid approximation. In
order to gauge the usefulness of the new criterion in this context, we
need to address two specific questions: First, how sharply does the
antesonic condition resolve differences between optimistic and
pessimistic models in the accretion phase, and, second, can it
accurately capture the onset of the runaway and the development of an
explosion?  To this end, we plot the time evolution of $\max
(c_s^2/v_\mathrm{esc}^2)$ for our 2D models in
Figure~\ref{fig:pt_criterion} using two different evaluation methods,
viz.\ averaging $c_s$ over solid angle before taking the maximum
(Case~A), and directly computing the maximum of the ratio
$c_s^2/v_\mathrm{esc}^2$ in the post-shock region (Case~B). As the
global maximum is very sensitive to fluctuations and not necessarily
indicative for the overall conditions in the gain region, Case~A seems
more in keeping with the 1D analysis of \citet{pejcha_11}.

Interestingly, we do not observe any clear differences in $\max (
c_s^2/v_\mathrm{esc}^2)$ between the 2D models in Case~A prior to the
development of an explosion. After an initial growth from $0.1$ to
$0.17$, $\max (c_s^2/v_\mathrm{esc}^2)$ settles to a stable average
value of about $0.17$ in all our simulations of the $15 M_\odot$
progenitor. Furthermore, the evolution of $\max (
c_s^2/v_\mathrm{esc}^2)$ in 1D is also virtually indistinguishable
from 2D in the case of the non-exploding model S15 (not shown in
Figure~\ref{fig:pt_criterion} as the curves would almost
overlap).\footnote{High values of $\max (c_s^2/v_\mathrm{esc}^2)
  \gtrsim 0.15$ are found in all our 1D models. Such high values
  naturally follow from the jump conditions at the shock for the
  typical infall velocities in dynamical simulations, which are
  considerably higher than assumed by \citet{pejcha_11}. It is unclear
  to what extent the choice of the infall velocity or the omission of
  GR affects the findings of \citet{pejcha_11}.} If we compute the
absolute maximum of the velocity ratio of \citet{pejcha_11} between
the proto-neutron star and the shock (case~B), we see an increase to
values between $0.2$ and $0.35$ once violent convective and SASI
activity starts (as a result of the higher entropy in the convective
plumes and SASI lobes). However, a clear hierarchy between the
different $15 M_\odot$ models is again absent; model M15, which
appears to be the most pessimistic case according to the time-scale
ratio $\tau_\mathrm{adv}/\tau_\mathrm{heat}$, even reaches somewhat
higher values than model S15 on average.

On the other hand, the onset of the explosion is indeed correlated
with an increase in $\max (c_s^2/v_\mathrm{esc}^2)$ beyond $0.2$. In
Case~A this increase can be delayed significantly for a strongly
asymmetric explosion (as occurs in run G15). The absolute maximum of
$c_s^2/v_\mathrm{esc}^2$ appears to be a more robust indicator, but
the critical value would have to be quite different from that
advocated by \citet{pejcha_11}, namely,
\begin{equation}
\max \left( c_s^2/v_\mathrm{esc}^2 \right) \gtrsim 0.35,
\end{equation}
but the evidence from two explosion models and two non-exploding
models of one of the progenitors is far from conclusive. We find
no compelling link between this (incidental?) observation and the
arguments of \citet{pejcha_11}.

Thus, unlike the time-scale ratio
$\tau_\mathrm{adv}/\tau_\mathrm{heat}$, the antesonic condition does
not appear to be a very useful measure for distinguishing optimistic
and pessimistic models during the accretion phase, and is of limited
use for detecting the runaway situation that ultimately leads to the
explosion. There is a number of possible reasons for this apparent
conflict with the findings of \citet{pejcha_11}. Different from the
time-scale ratio $\tau_\mathrm{adv}/\tau_\mathrm{heat}$, the antesonic
condition does not make any explicit reference to the heating
conditions in the gain layer; it depends only on the pressure, density
and gravitational potential at a specific point in the accretion flow
(corresponding more or less to the gain radius), which may be crucial
for the existence of a stationary accretion flow in spherical
symmetry. What happens to such a local quantity beyond the critical
point in the dynamical flow of a multi-dimensional situation is
unclear, however, whereas $\tau_\mathrm{adv}/\tau_\mathrm{heat}$
contains global information directly related to the competing
influence of accretion and neutrino heating.

Moreover, convection and the SASI alter the structure of the
post-shock accretion flow, the heating conditions, and the propagation
of the shock considerably in a multi-dimensional setup. The
stratification of the gain layer is not only modified by turbulent
mixing; the deviation of the velocity field from that of a stationary
spherical accretion flow and the deformation of the shock also become
large enough to invalidate many of the assumptions the one-dimensional
model of \citet{pejcha_11} -- such as the time-independent,
spherically symmetric Euler equations (neglecting, e.g., the turbulent
pressure) and the simple form of the jump conditions at the shock
(neglecting the SASI). More extreme cases with simultaneous accretion
and shock expansion in a dipolar explosion (such as model G15) can
hardly be accommodated in a 1D setup at all. These complications
certainly have the potential to change the findings of
\citet{pejcha_11} both quantitatively and qualitatively, and to make
$\max ( c_s^2/v_\mathrm{esc}^2)$ an inferior runaway criterion for
multi-dimensional supernova simulations. As demonstrated in the
preceding sections, the time-scale ratio
$\tau_\mathrm{adv}/\tau_\mathrm{heat}$ is less affected by such
complications and therefore remains, in our opinion, a better
diagnostic quantity for the evolution towards the explosion.

\section{Summary and Conclusions}
\label{sec:summary}
In this paper, we have presented the first multi-dimensional (2D)
general relativistic simulations of supernova explosions with a
sophisticated neutrino transport treatment. Using the recently
introduced \textsc{Vertex-CoCoNuT} code \citep{mueller_10}, which is
based on a relativistic generalization of the variable Eddington
factor method for neutrino transport \citep{rampp_02} combined with
the ``ray-by-ray-plus'' approach for multi-dimensional problems
\citep{buras_06_a,bruenn_06}, we calculated the evolution of two
progenitors with $11.2 M_\odot$ \citep{woosley_02} and $15 M_\odot$
\citep{woosley_95}. For the $15 M_\odot$ case, we also performed
complementary simulations with a pseudo-Newtonian and a purely
Newtonian treatment of gravity, and with a simplified set of neutrino
interaction rates.  We have conducted a comprehensive analysis of the
shock propagation, the explosion dynamics, and the heating conditions
for these models. With regard to the major goals formulated in
Section~\ref{sec:intro}, our results can be summarized as follows:
\begin{enumerate}
\item Feasibility of multi-dimensional relativistic supernova
  simulations: For both the $11.2 M_\odot$ and the $15 M_\odot$
    progenitor, we obtain explosions in 2D and have been able to
  extend the simulations several hundreds of milliseconds into the
  post-bounce and the explosion phase, thus proving the
  relativistic \textsc{Vertex-CoCoNuT} code to be as robust and stable
  as its Newtonian counterpart \textsc{Prometheus-Vertex}.

\item {Qualitative verification of pseudo-Newtonian results:} For the
  $11.2 M_\odot$ progenitor, we confirm the explosions of
  \citet{buras_06_b} and \citet{marek_09}, and for the $15 M_\odot$
  progenitor, the evolution is very similar to different models
  computed with the pseudo-Newtonian \textsc{Prometheus-Vertex} code
  for the first $\sim 400 \ \mathrm{ms}$ (i.e.\ before the onset of the
  explosion in our GR model). As \textsc{Vertex-CoCoNuT} and
  \textsc{Prometheus-Vertex} employ completely independent hydro
  solvers and somewhat different routines for the moment equations,
  our results can be viewed as an important step towards code
  verification in the sense that \emph{different} codes based on a
  \emph{similar} physical model have been shown to produce
  \emph{similar} results. However, the different physics leads to
  important quantitative differences (see next item).

\item Role of general relativistic effects in the supernova problem:
  For the $15 M_\odot$ progenitor of \citet{woosley_95}, we performed
  a comparison of a relativistic, a purely Newtonian, and a
  pseudo-Newtonian simulation in order to determine the importance of
  GR effects during the post-bounce phase. We find significant
  differences on the order of several tens of percent in some
  explosion-relevant quantities with a tendency towards more
  optimistic heating conditions in the GR case. This diagnosis is
  confirmed by the fact that we observe an incipient explosion in the
  GR run at $\sim 400 \ \mathrm{ms}$, whereas the Newtonian and
  pseudo-Newtonian models fail to explode, at least until $\sim 500
  \ \mathrm{ms}$ after bounce. Different thermodynamic conditions in
  the neutrinospheric region have been identified as the most likely
  cause for the more optimistic evolution in GR. It is noteworthy that
  the pseudo-Newtonian approach -- despite a much better
  quantitative agreement with the GR model than for the Newtonian run
  -- still leads to a different outcome than the GR case. Interestingly,
  our GR result yields a similar (successful) late explosion as
  obtained by \citet{marek_09} or a pseudo-Newtonian $15 M_\odot$ model
  including rotation and a gravitational potential that has been found
  to somewhat overestimate GR corrections in spherical symmetry.

  Thus, contrary to claims in the literature \citep{nordhaus_10},
  general relativity may class as a ``tens of percent'' effect with a
  similar impact on the dynamics as dimensionality
  \citep{nordhaus_10,hanke_11,takiwaki_12} and the equation of state
  (Marek~et~al., in preparation). GR thus emerges as a key ingredient
  for accurate supernova simulations.

\item Influence of the neutrino physics input: Based on a $15 M_\odot$
  model computed with a simplified set of opacities, we also
  illustrated the sensitivity of the neutrino heating conditions to
  the neutrino interaction rates. Specifically, we found that the
  energy transfer from $\nu_{\mu/\tau}$ to the background medium in
  the neutrinospheric region by nucleon recoil has a beneficial
  effect, since it effectively provides a means of ``converting'' a
  few $10^\mathrm{51} \ \mathrm{erg} \ \mathrm{s}^{-1}$ from
  $\nu_{\mu/\tau}$ into $\nu_e$ and $\bar{\nu}_e$, which can then
  actively contribute to the heating in the gain layer. Without the
  improved opacities, we observe a considerable delay of the explosion
  in the $15 M_\odot$ case for at least another $>50 - 100\ \mathrm{ms}$
  compared to the onset of the explosion in the baseline model G15 at
  $\sim 400 \ \mathrm{ms}$. Our findings further substantiate claims
  \citep{bruenn_09} about the importance of state-of-the-art neutrino
  physics input for supernova simulations, although the effect appears
  to be smaller than that of GR.

\end{enumerate}

On the whole, the inclusion of general relativity in supernova models
has turned out to be much more than a marginal improvement, even
compared to the effective potential approach which has hitherto been
used by some groups
\citep{rampp_02,marek_06,bruenn_09,scheidegger_08}. The purely
Newtonian approach is definitely ruled out as a viable basis for
quantitatively accurate models of core-collapse supernovae.  GR may
prove more important than expected for accurately capturing the
physics of neutrino-driven explosions if the $15 M_\odot$ progenitor
is any indication. In a subsequent publication, we will demonstrate
that GR also has a large impact on the observational signatures {such
  as neutrinos and gravitational waves from supernova cores}.

This adds another interesting facet to the supernova
problem and probably indicates that the explosion mechanism does not
hinge on a single dominating factor after all. General relativity,
neutrino microphysics, the equation of state (\citealp{marek_09}), and
dimensionality effects \citep{nordhaus_10,hanke_11,takiwaki_12} may
equally contribute to fill the missing parts of the current picture.

Such a situation naturally opens several avenues for future research
efforts. Above all, the potentially beneficial effects of GR need to
be confirmed by high-resolution studies (although improved angular
resolution will, in all probability, only lead to more robust
explosions in 2D, cp.\ \citealt{hanke_11}). In the long run, a
comprehensive and self-consistent approach to supernova modeling,
covering the essential aspects of the problem (such as neutrino
transport, general relativity, 3D hydrodynamics) with the help of
highly accurate numerical algorithms, will be indispensable for a firm
quantitative understanding of core-collapse supernova
explosions. \textsc{Vertex-CoCoNuT} provides a possible platform for
further developments in that direction, e.g.\ the extension of the code
to 3D, or the inclusion of more accurate multi-angle transport. On the
other hand, a more thorough understanding of the individual components
of the supernova problem (interplay of SASI and convection, critical
explosion conditions, nucleosynthesis, neutron star kicks and spins,
etc.) is equally important, and this is no less true for the effects
of general relativity discussed in this paper. Having confined
ourselves to two progenitors with $11.2 M_\odot$ and $15 M_\odot$,
which produce neutron stars of moderate compactness, we believe that
one of the essential tasks will be that of probing deeper into the
strong-field regime, where general relativity can be expected to play
an even greater role.  In our view, the case of more massive
progenitors, possibly on the verge to black hole formation, therefore
merits particular attention in the future.

\acknowledgements We thank A.~Bauswein for providing us with
Figure~\ref{fig:ns_mr} and L.~H\"udepohl for
data from an up-to-date 1D run obtained with the effective potential
approximation. This work was supported by the Deutsche
Forschungsgemeinschaft through the Transregional Collaborative
Research Centers SFB/TR 27 ``Neutrinos and Beyond'' and SFB/TR 7
``Gravitational Wave Astronomy'' and the Cluster of Excellence EXC 153
``Origin and Structure of the Universe''
(http://www.universe-cluster.de). The computations were performed on
the IBM p690 and the SGI Altix 3700 of the Computer Center Garching
(RZG), the NEC SX-8 at the HLRS in Stuttgart, and on the JUMP and
JUROPA systems at the John von Neumann Institute for Computing (NIC)
in J\"ulich.

\end{document}